\documentclass[12pt, a4page]{article}
\usepackage[utf8]{inputenc}
\usepackage{tikz}
\usetikzlibrary {arrows.meta,bending,positioning}
\usepackage[T1]{fontenc}
\usepackage[margin=1.25in]{geometry}
\usepackage{natbib}
\setcitestyle{authoryear,open={(},close={)}}
\usepackage{titlesec}
\titleformat*{\section}{\large\bfseries}
\titleformat*{\subsection}{\normalsize\bfseries}
\titleformat*{\subsubsection}{\normalsize\bfseries}
\usepackage{bbm}
\newcommand{\ind}[1]{{\mathbbm{1}_{\left\{{#1}\right\}}}}

\usepackage{fancyhdr}
\usepackage{xcolor}
\usepackage{bbm}
\usepackage{verbatim}
\usepackage{enumitem}
\usepackage{dsfont}
\usepackage{graphicx}
\usepackage{colortbl}
\usepackage{pst-all}	
\usepackage{hhline}   

\usepackage{comment}
\usepackage{xcolor}
\usepackage{graphics}
\usepackage{graphicx}
\usepackage{color}
\usepackage{amsmath}
\usepackage{amsthm}
\usepackage{amsfonts}
\usepackage{geometry}
\usepackage[normalem]{ulem}
\usepackage{float}
\usepackage{dsfont} 

\usepackage{mathabx} 

\usepackage[breaklinks]{hyperref}
\hypersetup{colorlinks=true,linkcolor=blue,citecolor=blue}
\usepackage{dsfont} 
\usepackage{nicefrac}

\newtheorem{theorem}{Theorem}
\newtheorem{lemma}{Lemma}
\newtheorem{proposition}{Proposition}

\def\a{\alpha}
\def\d{\delta}
\def\s{\sigma}
\def\eps{\varepsilon}
\def\Var{\mathrm{Var}}
\def\given{\, | \,}

\def\E{\mathbb{E}}
\def\P{\mathbb{P}}

\def\Z{\mathbb{Z}}
\def\indic{\mathbbm{1}}
\def\F{\mathcal{F}}
\def\dd{\mathrm{d}}
\newcommand{\ubar}[1]{\text{\underline{$#1$}}}

\usepackage{xcolor}

\begin{document}

\title{Network and Timing Effects in Social Learning} \author{Wade Hann-Caruthers\thanks{Technion - Israel Institute of Technology. Email: whanncar@gmail.com.} \ \ \ Minghao Pan\thanks{Caltech. Email: mpan2@caltech.edu.} \ \ \  Omer Tamuz\thanks{Caltech. Email: tamuz@caltech.edu. Omer Tamuz was supported by  a BSF award (\#2018397) and a National Science Foundation CAREER award (DMS-1944153).}}

\date{}

\maketitle

\begin{abstract}
We consider a group of agents who can each take an irreversible costly action whose payoff depends on an unknown state. Agents learn about the state from private signals, as well as from past actions of their social network neighbors, which creates an incentive to postpone taking the action. We show that outcomes depend on network structure: on networks with a linear structure patient agents do not converge to the first-best action, while on regular directed tree networks they do.
\end{abstract}

\section{Introduction}

Social networks play an important role in shaping the decisions people make. We ask friends for advice and take cues from their decisions, and we also understand that our friends interact with their friends in the same manner, as do our friends' friends, and so on. In recent years, there has been a large and active body of research devoted to understanding how the structure of the social network affects large-scale patterns of decision-making \citep[for recent surveys, see e.g.][]{golub2017learning, bikhchandani2021information}. Nevertheless, we have only very sparse understanding of how the structure of the social network affects the flow of information between rational agents.

We study agents who each have to decide whether and when to \emph{adopt}, i.e., take an irreversible costly action. Examples of such actions include installing rooftop solar panels, undergoing corrective eye surgery, or, more mundanely, watching a new movie or reading a new book. 
We assume that there is uncertainty regarding whether it is beneficial to adopt, determined by an unknown state of nature which is either high or low. At time zero each agent receives a private signal regarding the state. In each subsequent time period agents can choose to adopt, and then they observe the decisions of their social network neighbors. If an agent has not yet adopted they have a chance to adopt again in the next time period, but the decision to adopt is irreversible.

Agents face a trade-off: should they use the information available to them now, or wait and see what their neighbors do? Because adoption is irreversible, agents prefer not to adopt if the information they have indicates that the state is likely to be low. But even an agent who believes that the state is likely to be high may still want to delay their decision, and only adopt if their neighbors do, since their actions reveal additional information about the state. 

The question we ask is about \emph{learning}: When agents are patient, do they take the optimal action in the long run, i.e., eventually adopt when the state is high and never adopt when it is low? That is, does information diffuse along the network? We show that the answer depends on the geometry of the social network. Some networks obstruct diffusion, so that  the probability that agents take the optimal action is low, regardless of how patient the agents are. On other networks information diffuses: as agents become more patient, their probability of taking the correct action approaches one. 

When the social network is linear, as when people live along a road in a low-population area, there is a bound on how likely agents are to make good choices, a bound which holds regardless of how patient the agents are.  This inability to efficiently aggregate information is driven by that fact that in linear networks, there are few channels for the flow of information, and rational behavior renders these channels fragile. Agents serve not only as information sources for other agents but also as information conduits, and the inefficiency is generated by their failure to internalize the impact their choices have on the communication of information through the network.

The importance of rational behavior in making the information channels fragile is underscored by the fact that this is an equilibrium phenomenon. As we show, it is possible on these networks, out of equilibrium, for all agents simultaneously to be arbitrarily likely to make the right decision in the long run.

In contrast to the linear network case, we show that on networks that are very rapidly expanding, as when people live in a city, patient agents are very likely to take the correct action. In particular, this holds on tree networks in which each agent observes at least two others. In the case of tree networks, independent information flows to each agent from multiple directions, and thus even if some channels are obstructed, there is sufficient redundancy for information to spread and for all agents to eventually choose correctly, with high probability.

This work leaves open many interesting questions. In particular, we are far from understanding what happens on general networks. We further discuss this in the conclusion.

\subsection{Related literature}
To the best of our knowledge, the literature of social learning on social networks assumes that the timing of agents' actions is determined exogenously. On the other hand, studies of endogenous action timing generally assume that agents all observe each other or observe some kind of summary statistic of the choices made so far, and so do not provide insights into the potential effects of network structure \citep[see, e.g.][]{chamley2004delays}. We study settings with both endogenous timing and social networks.

A sizeable literature has studied learning on social networks when there is no endogeneity in action timing. When each agent acts once at an exogenously specified time, \cite*{bikhchandani1992theory} show that when agents see the actions of all of their predecessors, society acts suboptimally with positive probability, and similar results have been shown when the observational network is random \citep*[see e.g.][]{acemoglu2011bayesian, lobel2015information, arieli2019multidimensional}. \cite{bala1998learning} show that when myopic agents that exhibit a certain kind of bounded rationality act repeatedly, bounded (out-)degree (together with an appropriate spread of priors) leads to learning, and \cite{mossel2015strategic} show that when rational agents act repeatedly, a kind of structural ``egalitarianism'' is sufficient for society to eventually make good decisions. Note that the line network is egalitarian, but does not allow aggregation of information in our model, and so the flow dynamics in \cite{mossel2015strategic} are very different than those we consider. 

\cite{molavi2018theory} study long-run outcomes in models of social learning on networks across a broad class of non-Bayesian updating rules. \cite{huang2024learning} study the rate of learning in a Bayesian setting and show that learning is slow on every (strongly-connected) social network.

There has also been work on learning in settings with endogenous action timing in which the agents all observe each others' actions. The closest work to ours in this setting is \cite{chamley2004delays}, who studies a very similar base game on the complete network. He provides a characterization of symmetric equilibria in this setting, as well as an interesting comparison to the results in \cite*{bikhchandani1992theory}. Several papers also explore sudden shared behavior that occurs in models of heterogeneous signal precision just after the most informed agent acts \citep[see e.g.][]{zhang1997strategic, grenadier1999information}. See Section 5 of \cite{bikhchandani2021information} for a survey of the literature on endogenous timing in social learning.

Within the operations research literature, the Bass model \citep{bass1969new}, a non-strategic, reduced-form, tractable model for the adoption of new products in a population, has been highly influential. It has also been extended to models including social networks \citep[see, e.g.,][]{fibich2010aggregate}. In these continuous time models agents adopt according to a process whose rate increases with the number of neighboring agents who have adopted; thus all agents eventually adopt. The tractability of this model allow for explicit solutions of adoption times and their comparison across different networks.

Within the economics literature, \cite{board2021learning} study a closely related model of adoption on a social network. In their model, as in ours, there is a state that determines if adopting is beneficial. Each agent activates at an exogenously determined random time, and observes whether their neighbors have adopted. If any neighbors have adopted, they also adopt. If not, they make a strategic decision of whether to inspect the product at a cost, which reveals the state. They then adopt if the state is high. Their model reduces the agents' problem to a single decision: whether to inspect at the time they activate, given that none of their neighbors adopted so far. As a result, their model is highly tractable, allowing them to study comparative statics in the parameters of the (random) network.

There are several key difference between our work and that of \cite{board2021learning}. In their model agents are short lived and only take an action at one exogenously determined time instant, and the key consideration is whether to acquire costly information. Instead, in our model the agents are long lived, information is free and exogenous, and the key strategic consideration is an endogenous timing decision. In their model agents only adopt in the high state, and they study how the time of adoption varies with network parameters. In contrast, we ask whether agents eventually take the correct action. Accordingly, the conclusions from the two models are of a different nature.

Another related paper, by the same authors, is \cite{board2024experimentation}, which studies a model of strategic experimentation on a social network. In every time period agents choose how much effort to exert, and observe their own successes, as well as those of their neighbors. There is a binary state of the world. In the high state, successes arrive according to a Poisson process with intensity proportional to effort. In the low state there are no successes. Once a success is observed, agents learn that the state is high, and hence continue to invest and generate successes, spreading the news that the state is high across the network.  A perhaps surprising result (Proposition 1) is that agents do not wait to exert effort, as one may expect from an environment with a free-riding incentive, but rather exert maximum effort from the start, stopping only if no successes are observed after some time. Thus, in contrast to our model, agents do not exercise strategic timing. 

\section{Model}

We consider a set of agents, $N$, which may be finite or countably infinite. There is a state $\theta$, unknown to the agents, which is equally likely to be high ($H$) or low ($L$). Each agent $i$ receives a private signal $s_i$ about the state. We assume that the agents' signals are i.i.d.\ conditional on the state, and that they are bounded; that is, we assume the induced posterior belief $\pi_i = \P[\theta=H \given s_i]$ is almost surely in $[a, b]$ for some $0 < a < b < 1$. We further assume that the signals are informative, i.e., the distribution of $p_i$ is not a point mass at the prior $1/2$. A simple example to have in mind are symmetric binary signals,\footnote{For purely technical reasons, we prove Theorem~\ref{thm:degree} in the nonatomic case, but we expect this result to extend to more general signals.} i.e., $\P[s_i=\theta \given \theta]=q$ for some $q \in (1/2,1)$. 

There are infinitely many discrete periods $t = 0, 1, 2, \dots$. In each period each agent can choose to adopt ($A$) or not adopt ($N$), and adopting is irreversible so that once an agent has chosen to adopt they must choose to adopt in all subsequent time periods.  In a period in which an agent has not adopted, the flow payoff is zero. In periods in which the agent has adopted, the flow payoff is 1 in the high state, and -1 in the low state.\footnote{Our results generalize to the case that these two utilities are some constants $\overline{u} > 0 > \underline{u}$. Note that because adoption is irreversible, this model is equivalent to a model in which adoption can be chosen at most once, incurs a one time cost, and yields a one time payoff that depends on the state. } Agents discount at a common rate $\d \in (0,1)$.


There is a social network graph $G$ describing the relationships between the agents. The set of nodes of $G$ is the set of players, and $(i,j)$ is an edge if agent $i$ observes the actions of agent $j$. We denote by $N_i$ the set of agents $j$ that $i$ observes. A network undirected if $(i,j)$ is an edge whenever $(j,i)$ is an edge. 
We assume that $G$ is finite degree (i.e.\ each agent observes and is observed by only finitely many other agents). 
Our first main example is the bi-infinite, undirected \textit{line}.
The line  can be thought of as the one-dimensional social network where there is one agent at each integer location, and each agent observes the agents who are at a distance one from them. The second main example is the \textit{directed $d$-regular tree}, in which each agent observes $d\geq 2$ other agents and is observed by one other agent, and there are no cycles in the graph. Another prototypical example to have in mind is the complete network, which is the social network in which every agent observes every other agent. 

Agent $i$'s action at time $t$ is $a_t^i \in \{A,N\}$. In each time period, each agent $i$ observes the adoption decisions of her neighbors in the previous period. Thus, the information available to agent $i$ at time $t$ is their private signal, as well as the history $h_t^i = \{a_{t'}^j\}_{j \in N_i,t'<t}$ of actions taken by their neighbors in the previous periods. Accordingly, a pure strategy $\sigma_i$ for an agent is an adoption decision for each period $t$ based on her private signal and the history $h_t^i$, subject to the constraint that adoption is irreversible, i.e., that if $\sigma_i(s_i,h^i_t)=A$ then $\sigma_i(s_i,h^i_{t+1})=A$ whenever $h^i_{t+1}$ is an extension of $h^i_t$.  Equivalently, agent $i$'s (pure) strategy is a choice of the adoption time $\tau_i(s_i, h^i_\infty) = \min\{t \,:\, \sigma_i(s_i,h^i_t)=A\}$. We denote strategy profiles by $\tau = (\tau_i)_{i \in N}$.  

Given the flow payoffs, agent $i$'s discounted utility is $1 \cdot \d^{\tau_i}$ if the state is high, $-1\cdot\d^{\tau_i}$ if the state is low, and $0$ if $\tau_i=\infty$, i.e., if she never adopts. We assume that agents are discounted expected utility maximizers. We will study Bayes-Nash Equilibria of this game. 

\subsection{Diffusion}

It takes time for information to diffuse through the social network. In order for agents to be likely to adopt when the state is high and unlikely to adopt when the state is low, it must be the case that many agents wait a long time before adopting. Accordingly, we focus on what decisions agents have made after a long time. We will say that agent $i$ was \textit{eventually correct} if $\tau_i < \infty$ and $\theta = H$, or if $\tau_i = \infty$ and $\theta = L$. That is, the event that agent $i$ was eventually correct is the event that either the state was high and they eventually adopted, or the state was low and they did not adopt.  We denote by
\begin{align*}
    p_i = \P[(\tau_i < \infty \text{ and }\theta=H) \text{ or } (\tau_i = \infty \text{ and }\theta=L)]
\end{align*}
the probability that agent $i$ was eventually correct. Studying the probability of eventual correctness is in line with much of the social learning literature, which has been  concerned with asymptotic outcomes such as herding since its inception. 

A related question is that of efficiency, which also takes into account the time at which the correct action was taken. As we shall see, the question of efficiency turns out to be uninteresting in this model, as for large networks, regardless of the discount factor, expected utilities are controlled by the strengths of the signals. In particular, when signals are symmetric binary and the equilibrium is symmetric, agents with high signals will mix in the first period, and so their expected utility will equal the expected utility of playing the game by themselves and adopting immediately, or never adopting. One can think of eventual correctness as efficiency from the point of view of a social planner who is much more patient than the agents.

On a finite social network, there is a trivial bound on how likely any agent is to be eventually correct, simply because the agents' decisions are made on the basis of finitely many private signals. However, for infinite social networks that do not have some form of  structural impediment to the flow of information, it should in principle be possible for agents to be arbitrarily likely to make the right decision in the long run. A first important question is to understand when such structural impediments exist. I.e., if we put incentives aside and prescribe the agents' actions, can they be eventually correct? As we show in Proposition~\ref{prop:no-physical-obstruction}, there are, in fact, no such obstructions: for infinite undirected networks such as the bi-infinite line, there is always a (non-equilibrium) protocol under which  every agent is highly likely to be eventually correct.

However, the agents' incentives do constrain behavior, and so it is not a priori clear if agents can also be eventually correct in equilibrium. The question we explore is: how does the structure of the social network impact the existence of equilibrium protocols under which agents are very likely to be eventually correct? We will show that for some infinite networks agents are not very likely to be eventually correct: formally, there is a  $\bar{p} < 1$ such that every agent's probability of being eventually correct in every equilibrium for every value of the discount factor $\d$ is at most $\bar{p}$. In contrast, we show that for other infinite networks, agents can be eventually correct: for every $p<1$ there exists a discount factor $\d$ and a corresponding equilibrium such that every agent is eventually correct with probability at least $p$.

\section{Results}


\subsection{Preliminary observations}

Our first observation is that in any equilibrium, agents follow a threshold strategy in the following sense: for each history $h^i_t$ observed by agent $i$ at time $t$, if agent $i$ has not adopted by period $t$, then there is a threshold $q$ such that if the private belief $\pi_i$ is strictly above $q$ the agent adopts, and if $\pi_i$ is strictly below $q$ they do not adopt. Note that if signals are nonatomic, the agent's strategy is fully determined by the choice of a threshold for each history; otherwise, the agent's strategy may involve mixing when $\pi_i = q$ (see Lemma~\ref{lem:threshold-strategies}). Another simple observation is that in every equilibrium, there are some agents who adopt with positive probability in period $t=0$; indeed, if in periods $t < T$, all agents adopt with probability $0$, then any agent who would have otherwise adopted in period $T$ is better off adopting in period $0$.

Beyond these two observations, it seems difficult to characterize equilibria on general networks. Our main results are obtained for tree networks, which are more tractable. A connected, undirected network is a tree if it has no cycles, i.e., if there is no sequence of three or more distinct agents $i_1,\ldots, i_n$ such that for $k=1,\ldots,n-1$, each agent $i_k$ observes $i_{k+1}$ and $i_n$ observes $i_1$. We say that directed network is a tree if the underlying undirected network (i.e., the network corresponding to the symmetrized observation relation) is a tree.

Tree networks enjoy a tractability advantage, since what an agent observes from each neighbor is independent, conditioned on the state and on the agent's own actions. In particular, for tree networks we are able to show that all equilibria follow a natural dynamic: in period $t=0$, some agents with high signals will adopt. In subsequent periods, there is no \emph{spontaneous adoption}: an agent will not adopt in period $t > 0$ if none of their neighbors adopted in period $t-1$. 
\begin{proposition}
    \label{prop:spontaneous}
    In every equilibrium $\tau$ on a tree network, in every period $t>0$, if $\tau_j \neq t-1$ for all $j \in N_i$, then $\tau_i \neq t$.
\end{proposition}
Intuitively, observing no new adoptions is evidence against the high state, and so should not induce adoption. Nevertheless, we show in Appendix~\ref{app:spontaneous} that there are networks (that are not trees) on which spontaneous adoption can happen in equilibrium.\footnote{We thank Ben Wincelberg for suggesting this example to us. It is related to the non-monotonicity in social learning on networks observed by \cite{acemoglu2011bayesian}; see their Appendix B.} 

This proposition implies that from the perspective of an outside observer who sees only the decisions the agents make, equilibrium outcomes look very much like outcomes in standard infection or diffusion models: some agents adopt immediately, and then the decision to adopt spreads  from the initial adopters until reaching agents who do not adopt. 

\subsection{The probability of being eventually correct}

When agents are impatient, there is a limit to how likely they can be to be eventually correct. 
\begin{proposition}
\label{prop:impatient}
    Fix a network $G$, a private signal distribution, and an agent $i$. For every $\bar{\d}<1$ there is an $\varepsilon>0$ such that for all discount factors $\d < \bar{\d}$ the probability $p_i$ of agent $i$ being eventually correct is at most $1 - \varepsilon$ in every equilibrium.
\end{proposition}
The idea behind the proof of this proposition is that when agents are impatient, they will not want to wait a long time for information to arrive, restricting the set of agents whose actions can conceivably influence their decision. Their decision is hence based on a bounded amount of information. Thus, to achieve high $p_i$, we must consider patient agents. 


We begin by exploring the line network in which the set of agents is identified with the integers $\mathbb{Z}$ and $(i,j)$ is an edge in the social network graph if $|i-j|=1$. Our first main result shows that this graph provides an example of a network in which agents are not very likely to be eventually correct. 

\begin{theorem}
\label{thm:line}
Let $G$ be the line network.  Then there is a $\bar{p}<1$, that depends on the private signal distribution, such that in every equilibrium, for every value of the discount factor $\d$, every agent's probability $p_i$ of being eventually correct is at most $\bar{p}$.
\end{theorem}

At the heart of Theorem~\ref{thm:line} is the fact that agents serve not only as sources of information for other agents, but also as media for the flow of information. They do not take into account the impact of their decisions on the information channels they are a part of, and because there are very few information channels, they create blockages which cannot be circumvented. 

\begin{figure}
\centering
\begin{tikzpicture}[<->,>=stealth,>={Latex[length=2mm,width=1mm]},sibling distance=2cm,level distance=2cm]

  \node[circle,draw,minimum size=1cm] {}
    child [grow=right] {node[circle,draw,minimum size=1cm] {}
      child [grow=right] {node[circle,draw,minimum size=1cm] {}
        child [grow=right] {node[circle,draw,minimum size=1cm] {}
          child [grow=right] {node[circle,draw,minimum size=1cm] {}}
        }
        child [grow=up] {node[circle,draw,minimum size=1cm] {}
          child [grow=right] {node[circle,draw,minimum size=1cm] {}
          child [grow=right] {node[circle,draw,minimum size=1cm] {}}}
        }
      }
    };
\end{tikzpicture}
\caption{A one-dimensional graph $G$ with $M(G)=2$. \label{fig:one-d}}
\end{figure}
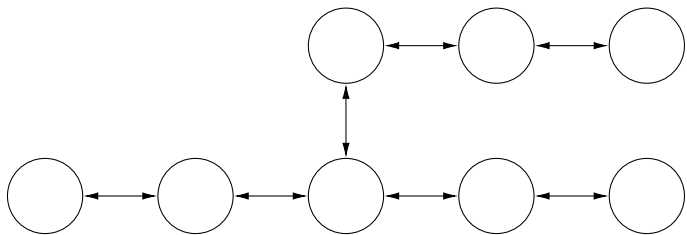

Our next result extends Theorem~\ref{thm:line} to a larger class of networks, which we call one-dimensional networks, and which also feature few available information channels. An undirected network is one-dimensional if it is a tree (i.e., has no cycles) and only finitely many agents have degree greater than two. These are graphs that can be formed by gluing together finitely many (finite or infinite) line segments; the minimum number of segments needed to form a one-dimensional graph $G$ is bounded above by
\begin{align*}
    M(G) = 1+\sum_i \max\{0,\deg(i)-2\},
\end{align*}
where $\deg(i)$ is the number of neighbors $i$ has (see Figure~\ref{fig:one-d}). 

For such networks, we obtain a bound on the probability of being eventually correct, which depends only on the private signal distribution and $M(G)$.

\begin{theorem}
\label{thm:one-dimensional}
Fix a private signal distribution and a number $m \geq 1$. Then there exists a  $\bar{p}<1$ such that for every one-dimensional network with $M(G)=m$, for every value of the discount factor $\d$, in every equilibrium,  every agent's probability $p_i$ of being eventually correct is at most $\bar{p}$.
\end{theorem}

In many classical learning models, information aggregation fails because only a small amount of information enters the system through the agents' actions. In contrast, in this model, a hypothetical outside observer, who could see the decisions made by all of the agents, would learn the state by observing just the actions taken in period $0$ (Proposition~\ref{prop:outside-observer-learns-immediately}). As we explore in the next section, the failure of information aggregation on one-dimensional networks is due to how well information is communicated across the network.

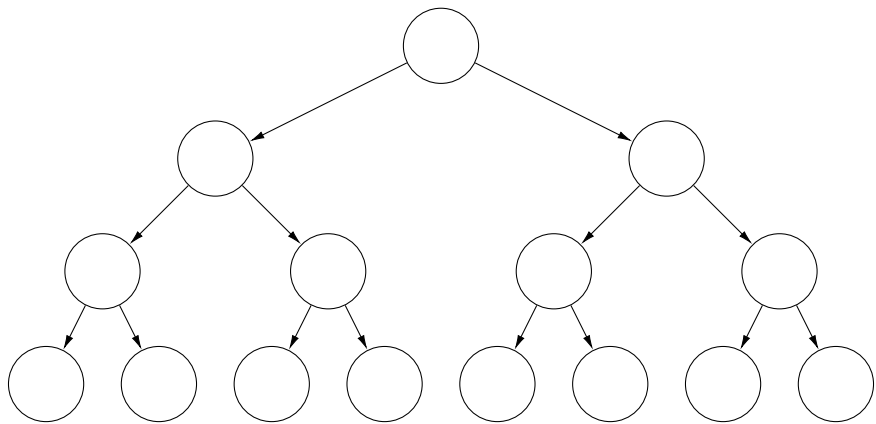
\begin{figure}
\centering
\begin{tikzpicture}[->,>=stealth,>={Latex[length=2mm,width=1mm]},level distance=1.5cm]
  \node[circle,draw,minimum size=1cm] {}
    [sibling distance=6cm]
    child {node[circle,draw,minimum size=1cm] {}
      [sibling distance=3cm]
      child {node[circle,draw,minimum size=1cm] {}
        [sibling distance=1.5cm]
        child {node[circle,draw,minimum size=1cm] {}}
        child {node[circle,draw,minimum size=1cm] {}}
      }
      child {node[circle,draw,minimum size=1cm] {}
        [sibling distance=1.5cm]
        child {node[circle,draw,minimum size=1cm] {}}
        child {node[circle,draw,minimum size=1cm] {}}
      }
    }
    child {node[circle,draw,minimum size=1cm] {}
      [sibling distance=3cm]
      child {node[circle,draw,minimum size=1cm] {}
        [sibling distance=1.5cm]
        child {node[circle,draw,minimum size=1cm] {}}
        child {node[circle,draw,minimum size=1cm] {}}
      }
      child {node[circle,draw,minimum size=1cm] {}
        [sibling distance=1.5cm]
        child {node[circle,draw,minimum size=1cm] {}}
        child {node[circle,draw,minimum size=1cm] {}}
      }
    };
\end{tikzpicture}
\caption{Layers $0$, $1$, $2$ and $3$ of a directed $2$-regular tree network. \label{fig:tree}}
\end{figure}

Next, we consider infinite, directed $d$-regular tree networks. In these networks, a root agent will observe $d \geq 2$ other agents, each of which will observe $d$ additional agents, etc. Thus agents are arranged in layers, with $d^\ell$ agents in layer $\ell \geq 0$, and each agent in layer $\ell$ observing a distinct group of $d$ agents in layer $\ell+1$. The graph is directed in the sense that agents do not observe the agents who observe them (see Figure~\ref{fig:tree}). For these networks, we sometimes say that agent $j$ is a child of agent $i$ if $i$ observes $j$. 

In these networks there are many channels for information to travel; observe, for example, that there are $d^k$ distinct (but not necessarily disjoint) observational paths of length $k$ arriving at each agent. We show that on these networks agents can be arbitrarily likely to be eventually correct. 

\begin{theorem}
\label{thm:trees}
Fix a private signal distribution. Let $G$ be a directed, $d$-regular tree. Then for every $\ubar{p} < 1$, for all sufficiently large discount factors $\delta < 1$, and for all symmetric equilibria, the probability $p_i$ of agent $i$ being eventually correct is at least $\ubar{p}$.
\end{theorem}
The idea behind the proof is the following. Since the equilibrium is symmetric, all agents have the same probability of being eventually correct. This stands in tension with the fact that agents can do better than their neighbors, by incorporating the signals contained in their actions; these signals are independent because of the tree structure. Of course, the information in the actions of neighbors can only be exploited one period later, resulting in some loss. But when agents are very patient, this loss only becomes significant once agents are very likely to be correct. As a result, in every symmetric equilibrium patient agents are very likely to eventually choose correctly.\footnote{Note that the stark contrast between trees and one-dimensional networks is not driven by the fact that the former are directed and the latter are undirected; the same result of Theorem~\ref{thm:line} holds for the directed line. Rather, information is well aggregated on the tree because there are many pathways for it to travel.  
}

Our final main result considers networks with high degree agents, i.e., agents who observe many others. Building on the work of \cite{chamley2004delays}, who considers large complete networks, we show that agents with many neighbors are likely to be eventually correct, assuming the discount factor is neither too high nor too low. To simplify the proof, we assume  that private beliefs are nonatomic, i.e., that $\pi_i$ does not equal any particular value with positive probability. 
\begin{theorem}
\label{thm:degree}
    Fix a private signal distribution that induces  nonatomic private beliefs. Then for each $d \geq 1$ there is a $\ubar{p}(d) < 1$ and a $\d(d) <1$ such that $\lim_d \ubar{p}(d) = 1$ and for $\delta(d)$ discounting agents, in every network every agent $i$ who observes $d$ others has probability $p_i \geq \ubar{p}(d)$ of being eventually correct in every equilibrium.
\end{theorem}
Note that the result is not stated for all agents patient enough, but rather for agents with some intermediate discount factor. The idea behind the proof is that for intermediate values of the discount factor, sufficiently many of an agent $i$'s $d$ neighbors will adopt in the first period, thus providing $i$ with a strong enough signal to make a decision in the second period which is likely to be correct. This theorem implies that on networks in which all agents observe many others, all agents are likely to be eventually correct. Note that the converse of this statement is not true: as Theorem~\ref{thm:trees} shows, it is possible for agents to be likely to be eventually correct even when the degrees are bounded.

\section{Equilibria on the line}

In this section we explain the ideas behind the proof of Theorem~\ref{thm:line}. The same ideas also underpin Theorem~\ref{thm:one-dimensional}.

Observe first that if an agent has a high probability of being eventually correct, then their adoption decisions must provide a strong signal about the state--indeed, an outside observer who learns only whether the agent eventually adopts or not must have a strong belief about the state with high probability. Accordingly, to prove Theorem~\ref{thm:line}, we show that the adoption decisions of a fixed agent do not provide a strong signal about the state.

By Proposition~\ref{prop:spontaneous},  agents may adopt at period $t=0$, and the remaining agents will only adopt after observing an adoption. Hence adoption spreads at unit speed from early adopters to their neighbors, then neighbors' neighbors etc, until an agent decides not to adopt, even though a neighbor adopted. Consequently, if the agent $i=0$ adopts at period $t$, it must be that one of the two agents who are located distance $t$ from agent $0$ (say agent $t$) adopted at period $0$, and that each of the agents between $1,\ldots,t-1$ adopted in turn after seeing one of their neighbors adopt.

This provides a picture of the information agent $i=0$  receives by observing their neighbor $j=1$: when agent $1$ adopts at time $t-1$, agent $0$ can infer the exact adoption decisions of agents $1$ through $t$. This implies that agent $0$ learns that agent $t$ had a high enough private belief to adopt at $t=0$, and that agents $1,\ldots, t-1$  had private beliefs that (i) were not high enough for them to adopt at time $t=0$, and (ii) were high enough to adopt upon seeing a neighbor adopt. I.e., the private beliefs $(\pi_1,\ldots,\pi_t)$ of agents $1,\ldots, t$ are in $\prod_{i=1}^t[\underline{a}_i,\overline{a}_i]$ for some bounds $\underline{a}_i,\overline{a}_i$. 

As we now explain, the information content in such an observation cannot be high. This is formalized in the following proposition that proves that this holds in a very general setting. The Kullback-Leibler divergence $D_{\mathrm{KL}}$ is a standard way of quantifying how different two distributions are. Given a binary state $\theta \in \{H,L\}$ and a signal with conditional distributions $\nu_H,\nu_L$, one can quantify the informativeness of the signal by $D_{\mathrm{KL}}(\nu_H|\nu_L)$; see \cite{pomatto2023cost} for a recent axiomatization of this measure.
\begin{lemma}
    \label{lemma:product-signal}
    Let $S \in \{0,1\}$ be a uniform binary random variable, let $(B_1,\ldots,B_n)$ be conditionally (on $S$) independent binary random variables, each taking values in $\{0,1\}$, and let $B = \prod_i{B_i}$ be the indicator of the event that $B_i=1$ for all $i$. Denote by $\nu_s$ the distribution of $B$ conditioned on $S=s$.
    
    Suppose there exists an $\eps>0$ such that $\P[S = 1 \given B_i = 1] \in [\eps,1-\eps]$. Then 
    \begin{align*}
        D_{\mathrm{KL}}(\nu_1 | \nu_0) \leq 2 \cdot \frac{\log(\eps)}{\log(1-\eps)}.
    \end{align*}
\end{lemma}
To gain some intuition for this result, suppose that the $B_i$ are symmetric, with $\P[B_i=S|S] = r \in (1/2,1)$. Then, when $n$ is low the event $B=1$ is not very informative, because it only aggregates a few signals. When $n$ is high this event is very informative, but it also happens very rarely, and so the $B$ does not contain much information about $S$ in the Kullback-Leibler sense.

Lemma~\ref{lemma:product-signal} provides a bound for the informativeness of a product event in the space of conditionally independent, uniformly bounded signals. Importantly, this bound is independent of the number of signals $n$; indeed, it applies even for $n=\infty$. The statement is symmetric with respect to the labels of events and signals, so the same result would apply if $B$ was the indicator of the event $\{B_1=b_1,\ldots,B_n=b_n\}$ for any $b \in \{0,1\}^n$. This result may have applications beyond our model. For examples, if $n$ voters have conditionally independent (but perhaps not identical) private signals about a state that determine their vote, then observing whether a particular voting profile was selected is a signal of bounded informativeness.

To apply this result to our setting, we let $B_i$ be the indicator of the event that $\pi_i \in [\underline{a}_i,\overline{a}_i]$, and let $S$ be the state $\theta$. When  agent $0$ observes that agent $1$ adopts at time $t$ they observe $B$, which, by the lemma, delivers only a bounded amount of information---independent of $t$---about the state.

Lemma~\ref{lemma:product-signal} hence implies that the observation of neighbors on the line provides a signal whose informativeness is bounded, independently of time. Thus, regardless of how patient agents are and how long they may wait to decide, adoption decisions cannot be very likely to be correct. In fact, the bound given in Lemma~\ref{lemma:product-signal} translates directly to a bound on agent $i$'s probability $p_i$ of being eventually correct in any equilibrium, a bound which depends only on the maximum private signal strength, yielding a proof of Theorem~\ref{thm:line}. The proof of Theorem~\ref{thm:one-dimensional} for one-dimensional networks makes similar use of Lemma~\ref{lemma:product-signal} and also allows us to provide a quantitative bound on $p_i$ which depends only on the maximum signal strength and $M(G)$.





\section{Equilibria on regular directed trees}

In this section we describe why agents are likely to be eventually correct on directed $d$-regular trees, for $d \geq 2$. We fix $d=2$, as the rest of the cases follow from this one.  

To gain some initial intuition, fix the discount factor $\d$, and consider a network with just two layers: a root agent and two children. The children observe no-one, and so in equilibrium they decide immediately whether or not to adopt, based on their private signals, and either adopt at period $t=0$ or never adopt. The root agent can adopt at period $t=0$ based on its private signal, or else wait a period, and decide at period $t=1$ after seeing the children's actions in the previous period. Since no more information will arrive after this, the root no incentive to wait longer, and so in equilibrium it has no incentive to adopt in any period $t>1$. 

Clearly, the root can choose at period $t=0$ using the same strategy as the children, and hence achieve the same expected discounted utility as them. In some cases, however, the root can do strictly better than the children. Suppose that private signals are symmetric and binary, i.e., $\P[s_i=\theta\given \theta] = r \in (1/2,1)$. Then observing the children's period $0$ actions reveals their private signals, and so if the root agent waits for period $t=1$, its action can be informed by three signals rather than one. If it does choose to wait, it is optimal for it to choose an action that matches the majority of these three signals. Equivalently, the root follows the children if they agree, and follows its own private signals when they disagree.

Perhaps counter-intuitively, for a given discount factor $\delta$, the root agent will wait to see the children's actions as long as the precision of the symmetric binary signals $r$ is \emph{small} enough. This is because the marginal benefit of two additional signals is large when signals are weak, and small when signals are strong. For example, if the precision is $r=99\%$, then following the majority of the three signals yields  probability of $99.97\%$ of choosing correctly, which translates to only a $2\%$ increase in utility, and hence is only beneficial for discount factors above $\delta = 0.98$. However, if $r=51\%$, then following the majority yields a $50\%$ increase in utility, making it beneficial already at low discount factors. As we shall see, this phenomenon is at the heart of our proof of Theorem~\ref{thm:trees}.

When signals are not symmetric, it might not be optimal for the root to follow the majority signal, even if it does decide to wait for the children's actions. For example, if positive signals are much stronger and more rare than negative ones, it will be optimal to adopt  if any of the three signals are positive.  Nevertheless, as our results below imply, if the children have low enough probability of choosing correctly, then there will be some strategy that makes it worthwhile for the root to wait one period and achieve strictly higher expected discounted utility than its children.

\medskip

Keeping this simplified model in mind, we return to our setting, with $\d<1$, an infinite $2$-regular directed tree, and a symmetric equilibrium. In this setting too, agents have the option of adopting if a majority of their children have adopted, adopting if any have adopted, etc. Just like in the setting above, we show that agents can do better than their children, if their children's ex-ante probability of being eventually correct is not too high. But this is impossible, since in a symmetric equilibrium all agents have the same probability of being eventually correct. Hence, if agents have low probability of being eventually correct, they cannot be in a symmetric equilibrium.

To prove this theorem we need to show that when the root's children employ the same strategy, and when they are not too likely to be eventually correct,  the root can achieve higher expected discounted utility than the children. It thus suffices to study the root's decision problem, fixing the children's strategies. We do this in the setting of an auxiliary model of the root's decision problem.

In this auxiliary model time is continuous and takes values in $[0,1]$. We embed the root's original decision problem into continuous time by mapping the discrete period $t$ to $1-\delta^t$, and discounting $t \in [0,1]$ by $1-t$. The advantage is that the root's decision problem is now a part of a larger---but compact---set of continuous time models. For this larger class we can use compactness to show that the root can do uniformly better than the children, if their probability of being eventually correct is not too high.

Formally, we consider the following auxiliary model. As in our main model, there is a binary state $\theta$ distributed uniformly over $\{H,L\}$. However, in contrast to our model, time is continuous and takes values in $[0,1]$. There are exactly three agents: a root agent $0$ and two children, $1$ and $2$. The children $1$ and $2$ adopt at times $\tau_1,\tau_2 \in [0,1]$, respectively, where the joint distribution of $(\tau_1,\theta)$ and $(\tau_2,\theta)$ are identical, and fixed exogenously. That is, we think of the children as employing a fixed strategy, and only consider the root's decision problem.

The root observes a private signal $s_0$ which is informative and bounded (as in our main model), and in addition observes at time $t$ which of the children (if any) have adopted so far. Time is discounted by $1-t$, so that an agent who adopts at time $\tau_i$ receives utility $1-\tau_i$ if $\theta=H$ and utility $-(1-\tau_i)$ if $\theta=L$.  Note that  $\tau_i=1$ yields utility zero regardless of the state; this will correspond to not adopting when we translate our model into this setting. 

We assume that the children's strategies $\tau_1,\tau_2$ satisfy the following condition: for every measurable $T \subset [0,1)$ it holds that $\P[\tau_i \in T\given \theta=H] \geq \P[\tau_i \in T\given \theta=L]$. This is a rationality assumption that captures the fact that in equilibrium agents are more likely to adopt in the high state than in the low state, a fact which will indeed hold when we translate our model into this setting.

The root's decision problem is to choose $\tau_0 \in [0,1]$ given the information available to it (the private signal and if and when the children have adopted), with the aim of maximizing its expected discounter utility. Formally, $\tau_0$  can be any stopping time with respect to the filtration $(\mathcal{F}_t)_{t \in [0,1]}$, where 
\begin{align*}
    \mathcal{F}_t = \sigma(s_0, (\ind{\tau_1 \leq s})_{s \leq t},(\ind{\tau_2 \leq s})_{s \leq t}).
\end{align*}
In particular, the root can choose (say) $\tau_0=\tau_1$, so that it can imitate a child immediately (rather than waiting one period, as in the discrete time model). 

We consider two families of strategies that the root can use to achieve this, $a_{1,r}$ and $a_{2,r}$, each parameterized by $r \in [0,1]$. As above, $\pi_0 = \P[\theta=H\given s_0]$ is the root's private belief. 
\begin{itemize}
    \item  $a_{1,r}$: The root never adopts in $[0,r)$. If child $1$ adopts by time $[0,r]$, then the root follows child $2$ starting at $t=r$ (if both children adopt by time $r$, the root adopts at time $r$). Otherwise, the root adopts when child $1$ adopts. I.e., 
    \begin{align*}
        \tau_0 = \begin{cases}
                \tau_1 &\text{if } \tau_1 > r\\
                \max\{\tau_2,r\} &\text{if } \tau_1 \leq r
        \end{cases}
    \end{align*}
    Thinking of $r$ as a relatively early time, this family of strategies entails following the first child unless it adopts early, in which case the second child is followed; the private signal is completely disregarded. This strategy is useful when a child is likely to be correct, unless it acts early, or when it is likely optimal to adopt if both children have adopted.
    
    \item $a_{2,r}$: If child $1$ has not adopted by time $r$, child $2$ adopts by time $r$,  and $\pi_0>1/2$, then the root adopts at time $r$. Otherwise, the root adopts when child $1$ adopts. I.e.,
    \begin{align*}
        \tau_0 = \begin{cases}
            r&\text{if } \tau_1  >r, \tau_2 \leq r, \pi_0>1/2\\
            \tau_1&\text{otherwise}.
        \end{cases}
    \end{align*}
    Here, the root follows its own signal at time $r$ when the children disagree about adopting, much like taking the majority of the signals in the two period model discussed above. 

\end{itemize}

Given a strategy of the chidlren, we show that at least one of the strategies in one of these families yields higher expected utility for the root, as compared to the children. Compactness allows us to bound this improvement from below, uniformly over all strategies of the children. This implies that, when translated back to our main model, following one of these strategies yields strictly higher expected utility for the root, even accounting for the one period delay when mimicking, assuming agents are patient enough.

\section{Conclusion}


In this paper, we have shown that network structure can have a profound influence on aggregate outcomes in adoption models when timing is endogenous. In particular, in some networks the agents' strategic behavior leads them to obstruct the flow of information, which results in limited information aggregation, while in others there is no such obstruction.

The model studied in this paper is rich, and we expect that we have only scratched the surface. In particular, we are not able to say much about networks that are not trees; their analysis is complicated by the fact that on these networks we cannot exclude spontaneous adoption (see Appendix~\ref{app:spontaneous}).   As a stark example, consider the ladder network: two copies of the line networks in which agents are also connected to their corresponding copy. Even though this network is very similar to the line network, we do not know if Theorem~\ref{thm:line} applies to it.  Likewise, we do not know if Theorem~\ref{thm:trees} holds for undirected regular trees, or more generally for graphs that have similar geometric properties. We leave these questions for future research.


\newpage

\bibliography{refs}

\begin{thebibliography}{18}
\providecommand{\natexlab}[1]{#1}
\providecommand{\url}[1]{\texttt{#1}}
\expandafter\ifx\csname urlstyle\endcsname\relax
  \providecommand{\doi}[1]{doi: #1}\else
  \providecommand{\doi}{doi: \begingroup \urlstyle{rm}\Url}\fi

\bibitem[Acemoglu et~al.(2011)Acemoglu, Dahleh, Lobel, and
  Ozdaglar]{acemoglu2011bayesian}
D.~Acemoglu, M.~A. Dahleh, I.~Lobel, and A.~Ozdaglar.
\newblock Bayesian learning in social networks.
\newblock \emph{The Review of Economic Studies}, 78\penalty0 (4):\penalty0
  1201--1236, 2011.

\bibitem[Arieli and Mueller-Frank(2019)]{arieli2019multidimensional}
I.~Arieli and M.~Mueller-Frank.
\newblock Multidimensional social learning.
\newblock \emph{The Review of Economic Studies}, 86\penalty0 (3):\penalty0
  913--940, 2019.

\bibitem[Bala and Goyal(1998)]{bala1998learning}
V.~Bala and S.~Goyal.
\newblock Learning from neighbours.
\newblock \emph{The review of economic studies}, 65\penalty0 (3):\penalty0
  595--621, 1998.

\bibitem[Bass(1969)]{bass1969new}
F.~M. Bass.
\newblock A new product growth for model consumer durables.
\newblock \emph{Management science}, 15\penalty0 (5):\penalty0 215--227, 1969.

\bibitem[Bikhchandani et~al.(1992)Bikhchandani, Hirshleifer, and
  Welch]{bikhchandani1992theory}
S.~Bikhchandani, D.~Hirshleifer, and I.~Welch.
\newblock A theory of fads, fashion, custom, and cultural change as
  informational cascades.
\newblock \emph{Journal of political Economy}, 100\penalty0 (5):\penalty0
  992--1026, 1992.

\bibitem[Bikhchandani et~al.(2021)Bikhchandani, Hirshleifer, Tamuz, and
  Welch]{bikhchandani2021information}
S.~Bikhchandani, D.~Hirshleifer, O.~Tamuz, and I.~Welch.
\newblock Information cascades and social learning.
\newblock Technical report, National Bureau of Economic Research, 2021.

\bibitem[Board and Meyer-ter{-}Vehn(2021)]{board2021learning}
S.~Board and M.~Meyer-ter{-}Vehn.
\newblock Learning dynamics in social networks.
\newblock \emph{Econometrica}, 89\penalty0 (6):\penalty0 2601--2635, 2021.

\bibitem[Board and Meyer-ter Vehn(2024)]{board2024experimentation}
S.~Board and M.~Meyer-ter Vehn.
\newblock Experimentation in networks.
\newblock \emph{American Economic Review}, 114\penalty0 (9):\penalty0
  2940--2980, 2024.

\bibitem[Chamley(2004)]{chamley2004delays}
C.~Chamley.
\newblock Delays and equilibria with large and small information in social
  learning.
\newblock \emph{European Economic Review}, 48\penalty0 (3):\penalty0 477--501,
  2004.

\bibitem[Fibich and Gibori(2010)]{fibich2010aggregate}
G.~Fibich and R.~Gibori.
\newblock Aggregate diffusion dynamics in agent-based models with a spatial
  structure.
\newblock \emph{Operations Research}, 58\penalty0 (5):\penalty0 1450--1468,
  2010.

\bibitem[Golub and Sadler(2017)]{golub2017learning}
B.~Golub and E.~Sadler.
\newblock Learning in social networks.
\newblock \emph{Available at SSRN 2919146}, 2017.

\bibitem[Grenadier(1999)]{grenadier1999information}
S.~R. Grenadier.
\newblock Information revelation through option exercise.
\newblock \emph{The Review of Financial Studies}, 12\penalty0 (1):\penalty0
  95--129, 1999.

\bibitem[Huang et~al.(2024)Huang, Strack, and Tamuz]{huang2024learning}
W.~Huang, P.~Strack, and O.~Tamuz.
\newblock Learning in repeated interactions on networks.
\newblock \emph{Econometrica}, 92\penalty0 (1):\penalty0 1--27, 2024.

\bibitem[Lobel and Sadler(2015)]{lobel2015information}
I.~Lobel and E.~Sadler.
\newblock Information diffusion in networks through social learning.
\newblock \emph{Theoretical Economics}, 10\penalty0 (3):\penalty0 807--851,
  2015.

\bibitem[Molavi et~al.(2018)Molavi, Tahbaz-Salehi, and
  Jadbabaie]{molavi2018theory}
P.~Molavi, A.~Tahbaz-Salehi, and A.~Jadbabaie.
\newblock A theory of non-bayesian social learning.
\newblock \emph{Econometrica}, 86\penalty0 (2):\penalty0 445--490, 2018.

\bibitem[Mossel et~al.(2015)Mossel, Sly, and Tamuz]{mossel2015strategic}
E.~Mossel, A.~Sly, and O.~Tamuz.
\newblock Strategic learning and the topology of social networks.
\newblock \emph{Econometrica}, 83\penalty0 (5):\penalty0 1755--1794, 2015.

\bibitem[Pomatto et~al.(2023)Pomatto, Strack, and Tamuz]{pomatto2023cost}
L.~Pomatto, P.~Strack, and O.~Tamuz.
\newblock The cost of information: The case of constant marginal costs.
\newblock \emph{American Economic Review}, 113\penalty0 (5):\penalty0
  1360--1393, 2023.

\bibitem[Zhang(1997)]{zhang1997strategic}
J.~Zhang.
\newblock Strategic delay and the onset of investment cascades.
\newblock \emph{The RAND Journal of Economics}, pages 188--205, 1997.

\end{thebibliography}

\newpage

\appendix

\section{Preliminary observations}

In this section, we develop some features of agents' behavior in equilibrium that will be needed in the subsequent analysis. We begin with the simple observation that agents use history-dependent thresholding when deciding whether to adopt.
\begin{lemma}
\label{lem:threshold-strategies}
Suppose $\tau$ is an equilibrium. Then for every agent $i$ and history $h_t^i$ such that $a_{t'}^i = N$ for $t' < t$, there is a threshold $\ubar{\pi}$ such that $a_t^i = A$ if $\pi_i > \ubar{\pi}$ and $a_t^i = N$ if $\pi_i < \ubar{\pi}$.
\end{lemma}

Note in particular that if the distribution of posteriors at every time $t$ induced by $\F_t$ is nonatomic, then there is a unique optimal stopping rule.

\begin{proof}
Let $h_t^i$ be such a history, and let $\tilde{\tau}_i$ be any continuation strategy for agent $i$ in which the agent does not adopt in period $t$. We prove the claim by showing that if $\tilde{\tau}_i$ yields higher utility than adopting at period $t$ at private belief $\pi_i$, then it still does at lower private beliefs.

Denote $X = \indic_{\theta = H} - \indic_{\theta = L}$. Suppose that the agents observed a signal $s_i=s$, which induced a private belief $\pi_i(s) = \P[\theta=H \given s_i=s]$. Observe that the difference between agent $i$'s expected utility if she follows $\tilde{\tau}_i$ and if she instead adopts is $\E[X \cdot (\d^{\tilde{\tau}_i} - \d^t) \given h_t^i, s_i=s]$. Since $\d^{\tilde{\tau}_i} < \d^t$ with probability one, it follows that $\tilde{\tau}_i$ is a strict improvement over adopting at time $t$ if and only if
\begin{align*}
    \lefteqn{\pi_i(s) \cdot \P[h_t^i \given \theta = H] \cdot \E[\d^{\tilde{\tau}_i} - \d^t \given h_t^i, \theta = H]}\\
    &> (1 - \pi_i(s)) \cdot \P[h_t^i \given \theta = L] \cdot \E[\d^{\tilde{\tau}_i} - \d^t \given h_t^i, \theta = L].
\end{align*}
Since $\delta^{\tilde{\tau}_i} < \delta ^t$, this holds if and only if
\begin{align*}
    \frac{\pi_i(s)}{1-\pi_i(s)} < \frac{\P[h_t^i \given \theta = L] \cdot \E[\d^{\tilde{\tau}_i} - \d^t \given h_t^i, \theta = L]}{\P[h_t^i \given \theta = H] \cdot \E[\d^{\tilde{\tau}_i} - \d^t \given h_t^i, \theta = H]} \cdot
\end{align*}    
Hence, if adopting immediately is suboptimal given the signal $s$, it is also suboptimal given any signal $s'$ such that $\pi_i(s') \leq \pi_i(s)$. Hence, taking $\ubar{\pi}$ to be the lowest private belief such that if $\pi_i(s) = \ubar{\pi}$ then there is no continuation strategy which is a strict improvement over adopting immediately, the result follows.
\end{proof}

Next, we observe that since agents never adopt when the state is more likely to be low, it follows that agents are more likely to adopt at any time in the high state than in the low state.

\begin{lemma}
\label{lem:adoption-more-likely-in-high-state}
In any equilibrium $\tau$, $\P[\tau_i = t \given \theta = H] \geq \P[\tau_i = t \given \theta = L]$ for every agent $i$ and every $t \geq 0$.
\end{lemma}

\begin{proof}
Since never adopting gives utility $0$, it follows that for every history $h^i_t$ and almost every signal $s_i$ such that agent $i$ adopts at time $t$, 
\begin{align*}
    \P[\theta = H \given h^i_t, s_i] \geq \P[\theta = L \given h^i_t, s_i],
\end{align*}
so by the law of iterated expectations,
\begin{align*}
    \P[\theta = H \given \tau_i = t] \geq \P[\theta = L \given \tau_i = t].
\end{align*}
Thus,
\begin{align*}
    \frac{\P[\tau_i = t \given \theta = H]}{\P[\tau_i = t \given \theta = L]} = \frac{\P[\theta = H \given \tau_i = t]}{\P[\theta = L \given \tau_i = t]} \geq 1.
\end{align*}
\end{proof}

\section{Proof of Proposition~\ref{prop:impatient}}


\begin{proof}[Proof of Proposition~\ref{prop:impatient}]
If $G$ is finite, then the probability of being eventually correct is bounded by the probability of guessing the state correctly after observing all of the private signals, and so the result follows immediately. Hence, assume $G$ is infinite.

Denote by $u_0$ the expected utility of adopting at $t = 0$ if $\pi_i \geq \frac12$ and never adopting otherwise. Observe that agent $i$'s expected utility in equilibrium must be at least $u_0$.

Fix $T$ such that $\bar{\d}^T < u_0$, and let $q$ be the probability that agent $i$ adopts before time $T$ in the high state. Then agent $i$'s expected utility is at most $\frac12 \cdot (q + (1 - q) \cdot \d^T) \leq \frac12 \cdot (q + \d^T)$, so $\frac12 \cdot (q + \d^T) \geq u_0$. It follows that $q \geq 2 u_0 - \d^T \geq u_0$.

Let $m$ be the number of agents within a distance $T$ of agent $i$, and denote by $\rho$ the strongest possible belief from observing $m$ signals. Note that the probability that agent $i$ adopts before time $T$ in the low state is at least $\frac{1 - \rho}{\rho} \cdot q \geq \frac{1 - \rho}{\rho} \cdot u_0$, and so the probability of being eventually correct is at most $1 - \frac{1 - \rho}{\rho} \cdot u_0$.
\end{proof}

\section{Proof of Theorem~\ref{thm:degree}}

To prove Theorem~\ref{thm:degree}, we begin with a few observations. Denote by $\pi_i^t = \P[\theta = H \given h_t^i, s_i]$ agent $i$'s belief at time $t$. We first note that if $\pi_i^t \geq \frac{1}{2 - \d}$, then agent $i$ adopts at time $t$. This follows from the fact that agent $i$'s expected utility from adopting at time $t$ is $[\pi_i^t - (1 - \pi_i^t)] \cdot \d^t$, while her expected utility from not adopting is strictly bounded above by $\pi_i^t \cdot \d^{t+1}$. This is particularly useful because the threshold $\frac{1}{2-\d}$ is independent of $t$. Hence, if an agent's belief ever becomes strong enough, the agent must adopt in any equilibrium.

Now, define $\bar{\d}$ by $(2 - \bar{\d})^{-1} = b$, where we recall that $\text{Supp } \pi_i = [a, b]$. If $\d < \bar{\d}$, then $\pi_i \geq \frac{1}{2 - \d}$ with strictly positive probability. If $\d$ is much smaller than $\bar{\d}$, then the probability that agent $i$ adopts immediately in the low state cannot be close to $0$. It follows that if $\d < \bar{\d}$, then in order for the probability that agent $i$ is eventually correct to be close to $1$, $\d$ must be close to $\bar{\d}$. On the other hand, if $\d \geq \bar{\d}$, then even agents who get the strongest possible signal may not adopt immediately, and it is not clear what can be said in general about how agents behave in period $0$.

To prove Theorem~\ref{thm:degree}, we will show that if $d$ is large and $\d$ is close to $\bar{\d}$, then $p_i$ must be close to $1$. We begin by showing that if a large amount of information is contained in the period $0$ actions of agent $i$'s neighbors, then she must be likely to be eventually correct. More precisely, in terms of $\pi_i^1$, the belief that agent $i$ has after seeing the period $0$ actions of her neighbors, the  next lemma states that agent $i$ is very likely to be eventually correct if $\pi_i^1$ is very likely to be close to certainty.
\begin{lemma}
\label{lem:concentrated-implies-eventually-correct}
For all $\eps > 0$ sufficiently small, in any equilibrium $\tau$ with $\delta$-discounting, if 
\begin{align*}
    \P[\pi_i^1 \leq 1 - \eps \given s_i, \theta = H] \leq \eps~\text{ and }~\P[\pi_i^1 \geq \eps \given s_i, \theta = L] \leq \eps
\end{align*}
for almost all $s_i$, then the probability that agent $i$ is eventually correct is at least $(1 - 3 \eps) \cdot \P[\pi_i \leq \frac{1}{2 - \d} - 3 \eps]$.
\end{lemma}


\begin{proof}
Fix a signal realization $s_i$, and suppose that agent $i$ does not adopt in period $0$. Note first that if $1 - \eps > \frac{1}{2 - \d}$ and $\pi_i^1 \geq 1 - \eps$, then $\pi_i^1 > \frac{1}{2 - \d}$, and agent $i$ adopts in period $1$. Hence,
\begin{align*}
    \P[\tau_i = 1 \given s_i, \theta = H] \geq 1 - \eps \geq 1 - 3 \eps.
\end{align*}
Now, let $A$ be the event that there is some $t > 1$ such that $\pi_i^t \geq \frac12$. By Doob's martingale inequality,
\begin{align*}
     \P[A \given \pi_i^1 < \eps, \theta = L, s_i] \cdot \P[\theta = L \given \pi_i^1 < \eps, s_i] \leq \P[A \given \pi_i^1 < \eps, s_i] \leq 2\eps,
\end{align*}
so
\begin{align*}
    \P[A \given \pi_i^1 < \eps, \theta = L, s_i] \leq \frac{2 \eps}{1 - \eps} \cdot
\end{align*}
Thus,
\begin{align*}
    \P[\tau_i \neq \infty \given s_i, \theta = L] &\leq \P[\pi_i^1 \geq \eps \given s_i, \theta = L] + \P[\pi_i^1 < \eps \given s_i, \theta = L] \cdot \P[A \given \pi_i^1 < \eps, \theta = L, s_i]\\
    &\leq \eps + (1 - \eps) \cdot \frac{2 \eps}{1 - \eps}\\
    &= 3 \eps,
\end{align*}
so
\begin{align*}
    \P[\tau_i = \infty \given s_i, \theta = L] \geq 1 - 3 \eps.
\end{align*}
It follows that if agent $i$ does not adopt in period $0$, then the probability that she is eventually correct is at least $1 - 3\eps$.

Now, observe that if agent $i$ does not adopt in period $0$, then her expected utility is at least 
\begin{align*}
    \pi_i (1 - 3 \eps) \d - (1 - \pi_i) 3 \eps \d = \d (\pi_i - 3 \eps),
\end{align*}
and if she adopts in period $0$, then her expected utility is $\pi_i - (1 - \pi_i)$. For $\pi_i < \frac{1}{2 - \d} - 3 \eps$, it follows that
\begin{align*}
    \pi_i - (1 - \pi_i) < \d (\pi_i - 3 \eps),
\end{align*}
and hence agent $i$ does not adopt in period $0$. Thus, if $\pi_i < \frac{1}{2 - \d} - 3 \eps$, then she does not adopt in period $0$ and the probability that she is eventually correct is at least $1 - 3 \eps$, and the result follows.
\end{proof}

\begin{proof}[Proof of Theorem~\ref{thm:degree}]
Fix $\eta > 0$, and choose $\d < \bar{\d}$ such that $\P[\pi_i > \frac{1}{2 - \d}] < \frac12 \eta$. Denote by $S$ the neighbors of agent $i$, and observe that
\begin{align*}
    \log \frac{\pi_i^1}{1 - \pi_i^1} = \log \frac{\pi_i}{1 - \pi_i} + \chi_S,
\end{align*}
where $\chi_S = \sum_{j \in S} \log \frac{\P[\tau_j = 0 \given \theta = H]}{\P[\tau_j = 0 \given \theta = H]}$, so
\begin{align*}
    \chi_S - \log(\a) \leq \log \frac{\pi_i^1}{1 - \pi_i^1} \leq \chi_S + \log(\a)
\end{align*}
where $\a = \max(\frac{1-a}{a}, \frac{b}{1-b})$. Given $q \in (0, 1)$, define $q'$ such that
\begin{align*}
    \log \frac{q'}{1-q'} = \log \frac{q}{1-q} + \log(\a).
\end{align*}
It follows that
\begin{align*}
    \P[\pi_i^1 \geq q \given \theta = H] \geq \P[\chi_S \geq \log \frac{q'}{1-q'} \given \theta = H]
\end{align*}
and
\begin{align*}
    \P[\pi_i^1 \leq 1 - q \given \theta = L] \geq \P[\chi_S \leq -\log \frac{q'}{1 - q'} \given \theta = L],
\end{align*}

Now, for each $j \in S$,
\begin{align*}
    0 < \P[\pi_j \geq \frac{1}{2 - \d} \given \theta = L] \leq \P[\tau_j = 0 \given \theta = L] \leq \P[\pi_j \geq \frac12 \given \theta = L] < 1,
\end{align*}
\begin{align*}
    0 < \P[\pi_j \geq \frac{1}{2 - \d} \given \theta = H] \leq \P[\tau_j = 0 \given \theta = H] \leq \P[\pi_j \geq \frac12 \given \theta = H] < 1,
\end{align*}
and
\begin{align*}
    \frac{\P[\tau_j = 0 \given \theta = H]}{\P[\tau_j = 0 \given \theta = L]} \geq \frac{\P[\pi_j \geq \frac12 \given \theta = H]}{\P[\pi_j \geq \frac12 \given \theta = L]} > 1.
\end{align*}
Hence, by Proposition~\ref{prop:seeing-many-nbrs-informative}, for any $\eps \in (0, \frac12)$, there is an $m$ such that if $d \geq m$ then $\P[\pi_i^1 \geq 1 - \eps \given s_i, \theta = H] \geq 1 - \eps$ and $\P[\pi_i^1 \leq \eps \given s_i, \theta = L] \geq 1 - \eps$. 

Note that for all $\eps > 0$ sufficiently small, $\P[\pi_i > \frac{1}{2 - \d} - 3 \eps] < \eta$. Hence, by Lemma~\ref{lem:concentrated-implies-eventually-correct}, for every $\eps > 0$ there is a $m$ such that if $d \geq m$ then the probability that agent $i$ is eventually correct is at least $(1 - 3 \eps) \cdot \P[\pi_i \leq \frac{1}{2 - \d} - 3 \eps]$. Since this approaches $\P[\pi_i \leq \frac{1}{2 - \d}] > 1 - \frac12 \eta$, it follows that there is an $m$ such that if $d \geq m$ then the probability that agent $i$ is eventually correct is at least $1 - \eta$. 

In particular, it follows that for every $n$ there is a $\d_n$ and an $m_n$ such that with discount factor $\d_n$, if $d \geq m_n$ then the probability that an agent with $d$ neighbors is eventually correct is at least $1 - \frac1n$ in any equilibrium on any network. For every $d \geq m_2$, let $n$ be the largest such that $m_n \leq d$, let $\d(d) = \d_n$, and let $\ubar{p}(d) = 1 - \frac1n$. The result then follows.
\end{proof}

\section{No physical impediments}

\begin{proposition}
\label{prop:no-physical-obstruction}
Let $G$ be an infinite, connected, undirected network which contains a bi-infinite line. For any $\eps > 0$, there is a strategy profile such that every agent's probability of being eventually correct is at least $1 - \eps$.
\end{proposition}
The key observation underlying this result is that, because agents can choose not only whether to adopt but also a time at which to adopt, there is no bound on how much information can be communicated between neighbors. For example, if agent $i$ wanted to communicate a very good approximation of her belief to agent $i+1$, the agents could in principle agree on a protocol for when agent $i$ should adopt depending on her signal.

\begin{proof}
To begin, fix a bi-infinite path in $G$ and label the agents in the path by $\Z$ so that there is an edge between $i$ and $j$ if and only if $|i - j| = 1$. For each $i \in \Z$, define
\begin{align*}
    x_i = 
    \begin{cases}
        0& \text{if } \pi_i < 1/2\\
        1& \text{if } 1/2 \leq \pi_i
    \end{cases}
\end{align*}
and let $q(\theta) = \P[x_i = 1 \given \theta]$.

For $\eta \in (0, 1)$ and $k \geq 3$, we define the strategy profile $\s(\eta, k)$ for agents in $\Z$ as follows. Each agent $i \in \Z$ adopts independently in period $0$ with probability $\eta$. If agent $i$ does not adopt in period $0$, then agent $i$ uses the strategy
\begin{align*}
    \tau_i =
    \begin{cases}
        \tau_{i-1} + k + x_i, &\text{ if } \tau_{i-1} < (k-1) \cdot k\\
        k^2, &\text{ if } (k-1) \cdot k \leq \tau_{i-1} < k^2 \text{ and } \frac{\tau_{i-1} - (k-1) \cdot k + x_i}{k} > \frac{q(H) + q(L)}{2}\\
        \tau_{i-1} + 1, &\text{ if } \tau_{i-1} \geq k^2\\
        \infty, &\text{ otherwise}
    \end{cases}
\end{align*}

Observe first that for every agent not in $\Z$, their probability of being eventually correct is equal to the probability that the closest agent in $\Z$ is eventually correct. Moreover, by symmetry, the probability of being eventually correct is the same for all agents in $\Z$.

Note that by symmetry, the probability of being eventually correct is the same for all agents in $\Z$. Now, consider agent $0$, and let $j$ be the nearest neighbor to the left of agent $0$ who adopts immediately, $j := \max_{i \leq 0}{\tau_i = 0}$, and let $s = \frac{1}{k} \sum_{i = j+1}^{j+k}{x_i}$. Note that a sufficient condition for agent $0$ to be eventually correct is $j < -k$ and $s > \frac{q(H) + q(L)}{2}$ if $\theta = H$ and $s < \frac{q(H) + q(L)}{2}$ if $\theta = L$.

Fix $\eps \in (0, 1)$. Note that if $k$ is sufficiently large, $\P[s > \frac{q(H) + q(L)}{2} \given \theta = H] > \sqrt{1 - \eps}$ and $\P[s < \frac{q(H) + q(L)}{2} \given \theta = L] > \sqrt{1 - \eps}$. Similarly, if $\eta$ is sufficiently small, $\P[j < -k \given \theta] > \sqrt{1 - \eps}$. It follows that for all sufficiently large $k$ and all sufficiently small $\eta$, the probability that agent $0$ is eventually correct under $\s(\eta, k)$ is at least $1 - \eps$.

Finally, let $T$ be any spanning tree of $G$ which contains the chosen bi-infinite path $\Z$, and extend the strategy profile $\s(\eta, k)$ by setting the strategy of every agent not in $\Z$ to be adopt immediately if one of their neighbors in $T$ adopts and never adopt otherwise. Under this strategy profile, an agent not in $\Z$ adopts if and only if the closest agent in $\Z$ adopts. It follows that every agent has the same probability of being eventually correct, and so every agent's probability of being eventually correct is at least $1 - \eps$.
\end{proof}


One implication of Proposition~\ref{prop:no-physical-obstruction} is that, for patient agents, it is possible to approximate the first-best welfare outcome. Note that under any strategy profile, each agent's expected payoff is at most $1/2$, since each agent can get a payoff of at most $0$ in the low state and $1$ in the high state. Under the strategy profile in Proposition~\ref{prop:no-physical-obstruction}, the expected payoff to every agent nearly achieves this upper bound.



\section{No spontaneous adoption on trees}

Recall that a strategy $\s_i$ for agent $i$ satisfies \textit{no spontaneous adoption} if for every $t >0$ and every history $h^i_t$ in which none of her neighbors adopts in period $t-1$, agent $i$ does not adopt in period $t$. A a strategy profile satisfies no spontaneous adoption if every agent's strategy satisfies no spontaneous adoption. In this section we prove Proposition~\ref{prop:spontaneous}, which states that on tree networks there is a no spontaneous adoption in any equilibrium.



To this end if we first show that if $G$ is a tree, then seeing a neighbor adopt is always evidence that the state is high. For any strategy profile $\tau$, define the strategy profile $\ubar{\tau}^i$ to be the one under which all agents except $i$ use the same strategy, and agent $i$ uses the strategy under which they never adopt. Observe that in order for $\tau$ to be an equilibrium, $\tau_i$ must be an optimal stopping time with respect to $s_i$ and $(\ubar{\tau}^i_j)_{j \in N_i}$. 
\begin{lemma}
\label{lem:neighbor-invests-on-tree-good-news}
Suppose $G$ is a tree and agents follow an equilibrium strategy profile $\tau$. If agent $j$ observes agent $i$, then
\begin{align*}
    \P[\ubar{\tau}^i_j = t \given \theta = H] \geq \P[\ubar{\tau}^i_j = t \given \theta = L].
\end{align*}
\end{lemma}

\begin{proof}

First, since agent $j$'s adoption decision at time $0$ depends only on her signal, $\P[\ubar{\tau}^i_j = 0 \given \theta] = \P[\tau_j = 0 \given \theta]$, so by Lemma~\ref{lem:adoption-more-likely-in-high-state},
\begin{align*}
    \P[\ubar{\tau}^i_j = 0 \given \theta = H] \geq \P[\ubar{\tau}^i_j = 0 \given \theta = L].
\end{align*}
Now, suppose the claim holds for all $t' < t$. Observe that agent $j$'s action at time $t$ is a function of $s_j$ and $\indic(\ubar{\tau}_k^j = t')$ for $k \in N_j$ and $t' < t$. Moreover, since $G$ is a tree, $s_i$ and $(\ubar{\tau}_k^j)$ are conditionally independent random variables. Let $\mathcal{H} = (\indic(\ubar{\tau}_k^j = t'))_{k \in N_j \setminus i, t' < t}$, and let $h_t^j$ and $s_j$ be such under $\tau$, agent $i$ does not adopt before $t$ and agent $j$ adopts at $t$. Then
\begin{align*}
    1 &\leq \frac{\P[\theta = H \given \ubar{\tau}_i^j \geq t, \mathcal{H}, s_j]}{\P[\theta = L \given \ubar{\tau}_i^j \geq t, \mathcal{H}, s_j]}\\
    &= \frac{\P[\ubar{\tau}_i^j \geq t \given \theta = H] \cdot \P[\theta = H \given \mathcal{H}, s_j]}{\P[\ubar{\tau}_i^j \geq t \given \theta = L] \cdot \P[\theta = L \given \mathcal{H}, s_j]}\\
    &\leq \frac{\P[\theta = H \given \mathcal{H}, s_j]}{\P[\theta = L \given \mathcal{H}, s_j]} \cdot
\end{align*}
By the law of iterated expectations, it follows that
\begin{align*}
    \P[\theta = H \given \ubar{\tau}^i_j = t] \geq \P[\theta = L \given \ubar{\tau}^i_j = t],
\end{align*}
and the result then follows.
\end{proof}

An important corollary is that when $G$ is a tree, seeing more neighbors adopt is always stronger evidence that the state is high.
\begin{lemma}
\label{lem:on-trees-more-neighbors-stronger-evidence}
Suppose $G$ is a tree and agents follow an equilibrium strategy profile $\tau$. Fix a time $t$, let $s_i$ be a private signal and let $(h^i_1,h^i_2,\ldots)$ and $(\bar h^i_1,\bar h^i_2,\ldots)$ be histories such that $h^i_t = \bar h^i_{t}$, such that $i$ does not adopt before $t+1$ under either history, and such that the set of $i$'s neighbors who adopt in period $t$ in $h^i$ is a subset of the set of neighbors who adopt in period $t$ in $\bar h^i$. Then
\begin{align*}
    \P[\theta = H \given \bar h^i_{t+1}, \, s_i] \geq \P[\theta = H \given h^i_{t+1}, \, s_i].
\end{align*}
\end{lemma}

\begin{proof}
To begin, note that the claim holds if and only if
\begin{align*}
    \frac{\P[\theta = H \given \bar h^i_{t+1}, \, s_i]}{\P[\theta = L \given \bar h^i_{t+1}, \, s_i]} \geq \frac{\P[\theta = H \given h^i_{t+1}, \, s_i]}{\P[\theta = L \given h^i_{t+1}, \, s_i]}
\end{align*}
if and only if
\begin{align*}
    \frac{\P[\theta = H \given \bar h^i_{t+1}, \, s_i] / \P[\theta = L \given \bar h^i_{t+1}, \, s_i]}{\P[\theta = H \given h^i_{t+1}, \, s_i] / \P[\theta = L \given h^i_{t+1}, \, s_i]} \geq 1.
\end{align*}
Now,
\begin{align*}
    \frac{\P[\theta = H \given \bar h^i_{t+1}, \, s_i] / \P[\theta = L \given \bar h^i_{t+1}, \, s_i]}{\P[\theta = H \given h^i_{t+1}, \, s_i] / \P[\theta = L \given h^i_{t+1}, \, s_i]} &= \frac{\P[\bar h^i_{t+1} \given \theta = H] / \P[\bar h^i_{t+1} \given \theta = L]}{\P[h^i_{t+1} \given \theta = H] / \P[h^i_{t+1} \given \theta = L]}\\
    &= \prod_{j \in S}{\frac{\P[\ubar{\tau}^i_j = t \given \theta = H] / \P[\ubar{\tau}^i_j = t \given \theta = L]}{\P[\ubar{\tau}^i_j > t \given \theta = H] / \P[\ubar{\tau}^i_j > t \given \theta = L]}}
\end{align*}
where the last equality follows from conditional independence of the $\ubar{\tau}_j^i$. By Lemma~\ref{lem:neighbor-invests-on-tree-good-news}, every term in this product is at least $1$, and the result then follows.
\end{proof}

\begin{proof}[Proof of Proposition~\ref{prop:spontaneous}]
Suppose not. Then there is an equilibrium strategy profile $\tau$, an agent $i$, and a history $h_{t+1}^i$ such that none of $i$'s neighbors adopts in period $t$ but $i$ adopts in period $t+1$ with nonzero probability. 

Let $\bar h_{t+1}^i$ be a history such that $\bar h_t^i = h_t^i$, let $S$ be the set of $i$'s neighbors who adopt at time $t$ under $\bar h_{t+1}^i$, and let $T$ be the set of $i$'s neighbors who do not adopt by time $t$ under $\bar h_{t+1}^i$. Suppose $S \neq \emptyset$. Observe that among continuation strategies for agent $i$ after $h_{t+1}^i$ which depend only on the actions of agents in $T$, adopting in period $t+1$ must be optimal, since adopting in period $t+1$ is optimal among all continuation strategies. Now, by Lemma~\ref{lem:on-trees-more-neighbors-stronger-evidence},
\begin{align*}
    \P[\theta = H \given \bar h_{t+1}^i, s_i] \geq \P[\theta = H \given h_{t+1}^i, s_i],
\end{align*}
and so adopting in period $t+1$ must be optimal among continuation strategies after $\bar h_{t+1}^i$ which depend only on the actions of agents in $T$. Since all strategies take this form, it follows that it is optimal for agent $i$ to adopt in period $t+1$, contradiction.

Hence, it must be that agent $i$ adopts in period $t+1$ under every continuation of $h_t^i$. But in this case, switching to adopting in period $t$ improves agent $i$'s expected utility by a factor of $\d$, contradiction. Thus, the probability that agent $i$ adopts in period $t+1$ following any history where none of her neighbors adopts in period $t$ is zero.
\end{proof}

\section{Spontaneous adoption on non-trees}
\label{app:spontaneous}
Perhaps surprisingly, we show in this section that spontaneous adoption is possible for some social networks and some signal structures. Our construction is similar to other examples of non-monotonicity in social learning \citep[see][Appendix B]{acemoglu2011bayesian}. We thank Ben Wincelberg for suggesting it to us.

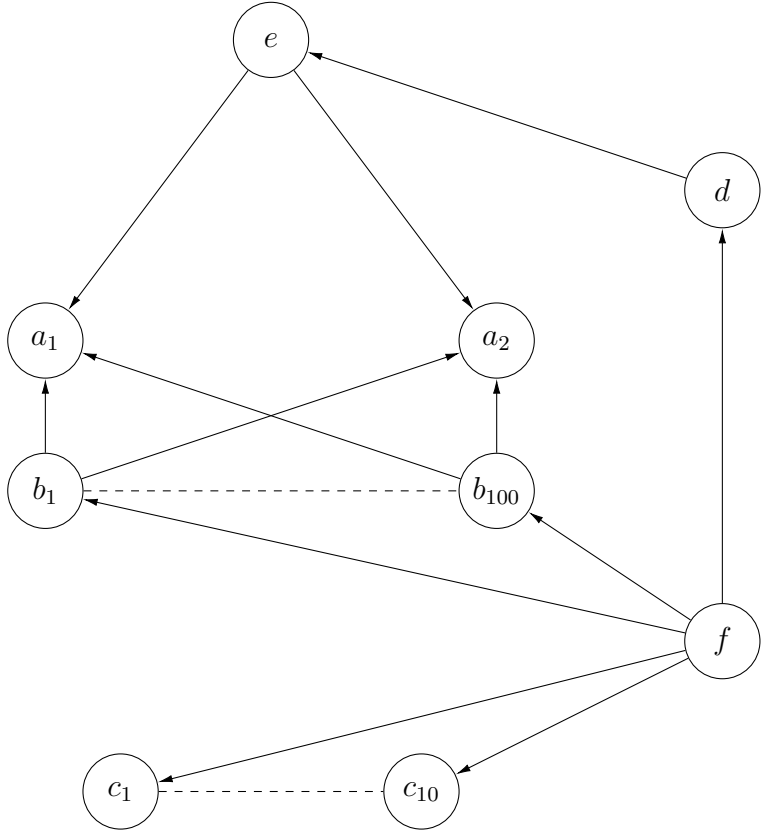
\begin{figure}
\centering
\begin{tikzpicture}[
    node/.style={circle, draw, minimum size=1cm, inner sep=0pt},
    obs/.style={->, >=stealth, >={Latex[length=2mm,width=1mm]}},
    midtext/.style={midway, font=\scriptsize, sloped}
]

\node[node] (a1) at (-2, 4) {$a_1$};
\node[node] (a2) at (4, 4) {$a_2$};
\node[node] (b1) at (-2, 2) {$b_1$};
\node[node] (b100) at (4, 2) {$b_{100}$};
\node[node] (c1) at (-1, -2) {$c_1$};
\node[node] (c10) at (3, -2) {$c_{10}$};
\node[node] (d) at (7, 6) {$d$};
\node[node] (e) at (1, 8) {$e$};
\node[node] (f) at (7, 0) {$f$};


\draw[obs] (b1) -- (a1);
\draw[obs] (b1) -- (a2);
\draw[obs] (b100) -- (a1);
\draw[obs] (b100) -- (a2);

\draw[obs] (d) -- (e);
\draw[obs] (e) -- (a1);
\draw[obs] (e) -- (a2);

\draw[obs] (f) -- (d);
\draw[obs] (f) -- (b1);
\draw[obs] (f) -- (b100);
\draw[obs] (f) -- (c1);
\draw[obs] (f) -- (c10);

\draw[dashed] (b1) -- (b100);
\draw[dashed] (c1) -- (c10);

\end{tikzpicture}
\caption{An example of a network with spontaneous adoption\label{fig:spontaneous}}
\end{figure}

Let the set of agents be $N = A \cup B \cup C \cup \{d, e, f\}$, where $A = \{a_1, a_2\}$, $B = \{b_1, \dots, b_{100}\}$, and $C = \{c_1, \dots, c_{10}\}$. The agents in $A$ and $C$ observe no-one, and the agents in $B$ observe the agents in $A$. Agent $d$ observes $e$, and agent $e$ observes the agents in $A$. Finally, agent $f$ observes agent $d$, the agents in $B$, and the agents in $C$. See Figure~\ref{fig:spontaneous}.

Let the private signals be binary, taking values in $\{L, H\}$, where $s_i$ matches the state with probability $q > \frac12$. Consider the event that $a_1$ and the agents in $B$ receive the signal $H$ and all other agents receive the signal $L$. 

Observe that since the agents in $A$ and $C$ do not observe others, $a_1$ adopts in period $0$, and $a_2$ and the agents in $C$ never adopt. Assuming $\d$ is close enough to $1$, agents in $B$ defer in period $0$, then adopt in period $1$. Agent $e$ never adopts, and so agent $d$ also never adopts.

Now, observe that agent $f$'s belief is below $\frac12$ in periods $0$ and $1$. After observing the agents in $B$ adopt in period $1$, agent $f$ infers that either $a_1$ and $a_2$ received signal $H$, or one received $H$, one received $L$, and all agents in $B$ received $H$. Hence, agent $f$'s likelihood ratio is
\begin{align*}
    \frac{1-q}{q} \cdot \left(\frac{1-q}{q}\right)^{10} \cdot \frac{q^2 + 2q(1-q) \cdot q^{100}}{(1-q)^2 + 2q(1-q) \cdot (1-q)^{100}} < 1,
\end{align*}
so agent $f$'s belief in period $2$ is below $\frac12$. Finally, after observing that agent $d$ does not adopt in period $2$, agent $f$ infers that agent $e$ received an $L$ signal and that exactly one of $a_1$ and $a_2$ received an $H$ signal. Hence, agent $f$ learns that there were strictly more $H$ signals than $L$ signals, and so agent $f$ adopts in period $3$, despite the fact that no agents adopted in period $2$.

\section{Proof of Lemma~\ref{lemma:product-signal}}

\def\g{\gamma}
\begin{lemma}
\label{lem:kl-one-bernoulli}
Let $S, B$ be binary random variables taking values in $\{0, 1\}$, and denote by $\nu_s$ be the distribution of $B$ conditioned on $S = s$. For $\g > 1$, if $\P[B = 1 \given S = 1]^{\g} \leq \P[B = 1 \given S = 0] \leq \P[B = 1 \given S = 1]^{\frac{1}{\g}}$, then
\begin{align*}
    D_{\mathrm{KL}}(\nu_1 | \nu_0) \leq 2\g.
\end{align*}
\end{lemma}

\begin{proof}
Recall that
\begin{align*}
    D_{\mathrm{KL}}(\nu_1 | \nu_0) = \sum_{b \in \{0, 1\}}{\P[B = b \given S = 1] \cdot \log\left(\frac{\P[B = b \given S = 1]}{\P[B = b \given S = 0]}\right)}.
\end{align*}
To bound the first term, observe that
\begin{align*}
    \frac{\P[B = 0 \given S = 1]}{\P[B = 0 \given S = 0]}
    = \frac{1 - \P[B = 1 \given S = 1]}{1 - \P[B = 1 \given S = 0]}
    \leq \frac{1 - \P[B = 1 \given S = 1]}{1 - \P[B = 1 \given S = 1]^{\frac{1}{\g}}} \leq \g,
\end{align*}
so
\begin{align*}
    \P[B = 0 \given S = 1] \cdot \log\left(\frac{\P[B = 0 \given S = 1]}{\P[B = 0 \given S = 0]}\right) \leq \log(\g)
\end{align*}
To bound the second term, observe that
\begin{align*}
    \log\left(\frac{\P[B = 1 \given S = 1]}{\P[B = 1 \given S = 0]}\right) \leq \log\left(\frac{\P[B = 1 \given S = 1]}{\P[B = 1 \given S = 1]^{\g}}\right) = (\g - 1) \cdot -\log(\P[B = 1 \given S = 1]),
\end{align*}
so
\begin{align*}
    \P[B = 1 \given S = 1] \cdot \log\left(\frac{\P[B = 1 \given S = 1]}{\P[B = 1 \given S = 0]}\right) &\leq (\g - 1) \cdot [- \P[B = 1 \given S = 1] \cdot \log(\P[B = 1 \given S = 1])]\\
    &\leq \g - 1.
\end{align*}
Hence,
\begin{align*}
    D_{\mathrm{KL}}(\nu_1 | \nu_0) \leq \log(\g) + \g - 1 \leq 2\g.
\end{align*}
\end{proof}

In order to apply Lemma~\ref{lem:kl-one-bernoulli} to the setting in Lemma~\ref{lemma:product-signal}, we will make use the following analytical observation.
\begin{lemma}
\label{lem:power-inequality}
Let $\a > 1$ and $x, y \in (0, 1)$. If $y \geq \frac{1}{\a} \cdot x$ and $1 - y \leq \a \cdot (1 - x)$,
then
\begin{align*}
    y \geq x^{\frac{\log(1+\a)}{\log(1 + \frac{1}{\a})}}.
\end{align*}
\end{lemma}

\begin{proof}
Let $f(t) = \max(\frac{1}{\a} t, 1 - \a + \a t)$ and $g(t) = t^{\frac{\log(1+\a)}{\log(1 + \frac{1}{\a})}}$. Observe that
\begin{align*}
    f(0) &= 0 = g(0)\\
    f(\frac{\a}{1+\a}) &= \frac{1}{1+\a} = g(\frac{\a}{1+\a})\\
    f(1) &= 1 = g(1).
\end{align*}
Since $\frac{\log(1+\a)}{\log(1 + \frac{1}{\a})} > 1$, $g$ is convex. Moreover, since $f$ is linear on $[0, \frac{\a}{1+\a}]$ and on $[\frac{\a}{1+\a}, 1]$, it follows that $f(t) \geq g(t)$ for $t \in [0, 1]$. Hence, $y \geq f(x) \geq g(x) = x^{\frac{\log(1+\a)}{\log(1 + \frac{1}{\a})}}$.
\end{proof}

\begin{proof}[Proof of Lemma~\ref{lemma:product-signal}]
Let $\a = \frac{1-\eps}{\eps}$. Observe that
\begin{align*}
    \a \cdot \P[B_i = 1 \given S = 1] \geq \P[B_i = 1 \given S = 0] \geq \frac{1}{\a} \cdot \P[B_i = 1 \given S = 1].
\end{align*}
It follows from Lemma~\ref{lem:power-inequality} that
\begin{align*}
    \P[B_i = 1 \given S = 1]^{\g} \leq \P[B_i = 1 \given S = 0] \leq \P[B_i = 1 \given S = 1]^{\frac{1}{\g}},
\end{align*}
where $\g = \frac{\log(1+\a)}{\log(1 + \frac{1}{\a})} = \frac{\log(\eps)}{\log(1-\eps)}$.

Now, since the $B_i$ are conditionally independent,
\begin{align*}
    \P[B = 1 \given S = s] = \prod{\P[B_i = 1 \given S = s]}
\end{align*}
for $s \in \{0, 1\}$. Hence,
\begin{align*}
    \P[B = 1 \given S = 1]^{\g} \leq \P[B = 1 \given S = 0] \leq \P[B = 1 \given S = 1]^{\frac{1}{\g}}.
\end{align*}
The result now follows from Lemma~\ref{lem:kl-one-bernoulli}.
\end{proof}

\def\I{\mathcal{I}}

\section{Proof of Theorem~\ref{thm:one-dimensional}}

In this section, we prove Theorem~\ref{thm:one-dimensional}. Note that Theorem~\ref{thm:line} is an immediate corollary.

We assume throughout this section that the private signals induce beliefs in $[\eps, 1 - \eps]$, and we denote $\a = \frac{1-\eps}{\eps}$. Note that for any subset $A \in [0, 1]$,
\begin{align*}
    \frac{1}{\a} \leq \frac{\P[s_i \in A \given \theta = H]}{\P[s_i \in A \given \theta = L]} \leq \a.
\end{align*}


Define
\begin{align*}
    \I(X) = \E[\log(\frac{\P[X \given \theta = H]}{\P[X \given \theta = L]}) \given \theta = H]
\end{align*}
and
\begin{align*}
    \I(X \given Y) = \E[\log(\frac{\P[X \given Y, \theta = H]}{\P[X \given Y, \theta = L]}) \given \theta = H].
\end{align*}
Note that $\I(X)$ is the Kullback-Leibler divergence between the distribution of $X$ in the high state and the low state.

We begin by using $\I(\tau_i)$ to provide a quantitative bound on the probability that agent $i$ is eventually correct.

\begin{lemma}
\label{lem:p-i-bound-using-I}
In any equilibrium, the probability $p_i$ that agent $i$ is eventually correct is at most $1 - \frac{1}{2} e^{-\I(\tau_i) - 3}$.
\end{lemma}

\begin{proof}
Let $\eta = 1 - p_i$. Observe that $\P[\tau_i < \infty \given \theta = L] \leq 2 \eta$ and $\P[\tau_i = \infty \given \theta = H] \leq 2 \eta$. Hence,
\begin{align*}
    \lefteqn{\I(\indic(\tau_i = \infty))}\\
    &= \P[\tau_i < \infty \given \theta = H] \cdot \log(\P[\tau_i < \infty \given \theta = H]) - \P[\tau_i < \infty \given \theta = H] \cdot \log(\P[\tau_i < \infty \given \theta = L])\\ 
    &\quad + \P[\tau_i = \infty \given \theta = H] \cdot \log(\P[\tau_i = \infty \given \theta = H]) - \P[\tau_i = \infty \given \theta = H] \cdot \log(\P[\tau_i = \infty \given \theta = L])\\
    &\geq -1 + (1 - 2 \eta) \cdot -\log(2\eta) + -1 + 0\\
    &\geq -3 - \log(2\eta),
\end{align*}
and thus,
\begin{align*}
    p_i = 1-\eta \leq 1-\frac12 \cdot e^{-\I(\indic(\tau_i = \infty))) - 3}.
\end{align*}
Since $\indic(\tau_i = \infty)$ is a function of $\tau_i$, $\I(\indic(\tau_i = \infty)) \leq \I(\tau_i)$, and the result then follows.
\end{proof}

In order to prove Theorem~\ref{thm:one-dimensional}, we will prove an auxiliary result on rooted, directed trees. Let $T$ be a rooted, directed tree with finitely many ends, and let $(s_i)_{i \in T}$ be conditionally iid with $\P[\theta = H \given s_i] \in [\eps, 1 - \eps]$. An adoption process associated to $T$ is a choice of adoption time $\tau_i$ associated to each vertex $i$ such that $\tau_i$ is a function of $s_i$ and $(\tau_j)_{N_T(i)}$, $\tau_i$ satisfies no spontaneous adoption, and for any $t$, $\P[\tau_i = t \given \theta = H] \geq \P[\tau_i = t \given \theta = L]$. We denote by $\tau_r$ the adoption time of the root.

\begin{proposition}
\label{prop:rt-one-end}
There is a $c_0$ such that for any rooted tree $T$ with one end and any associated adoption process, $\I(\tau_r) \leq c_0$.
\end{proposition}

\begin{proof}
Label each vertex in $T$ by the distance from the root, so that the root is labeled $0$, its child is labeled $1$, and so on.

Let $\tilde{\tau} = \min \{i \, : \, \tau_i = 0\}$, where $\tilde{\tau} = 0$ if $\tau_i \neq 0$ for all $i$. Observe that by no spontaneous adoption, $\tau_0 \in \{\tilde{\tau}, \infty\}$. It follows that $\tau$ is a function of $\tilde{\tau}$ and $\indic(\tau_0 = \tilde{\tau})$, so
\begin{align*}
    \I(\tau_0) &\leq \I(\tilde{\tau}, \indic(\tau_0 = \tilde{\tau}))\\
    &= \I(\tilde{\tau}) + \I(\indic(\tau_0 = \tilde{\tau}) \given \tilde{\tau}).
\end{align*}

We begin by bounding $\I(\tilde{\tau})$. Observe that because agents only adopt when believe the state is more likely to be high, $\P[\tau_i \neq 0 \given \theta = H] \leq \P[\tau_i \neq 0 \given \theta = L]$. Since $\tilde{\tau} = t$ if and only if $\tau_i \neq 0$ for $0 \leq i < t$ and $\tau_t = 0$, it follows that
\begin{align*}
    \frac{\P[\tilde{\tau} = t \given \theta = H]}{\P[\tilde{\tau} = t \given \theta = L]} \leq \frac{\P[\tau_t = 0 \given \theta = H]}{\P[\tau_t = 0 \given \theta = L]} \leq \a.
\end{align*}
Hence, $\I(\tilde{\tau}) \leq \log(\a)$.

Next, we bound $\I(\indic(\tau_0 = \tilde{\tau}) \given \tilde{\tau})$. For $0 \leq i < \tilde{\tau}$, define $B_i$ to be $1$ if agent $i$ receives a private signal for which they adopt at time $t-i$ if agent agent $i+1$ adopts at time $t - (i+1)$, and $0$ otherwise. Define $B = \prod{B_i}$. Observe that if $\tilde{\tau} \neq \infty$, then $\tau_0 = \tilde{\tau}$ if and only if $B = 1$. Moreover, note that
\begin{align*}
    \frac{\P[B_i = 1 \given \tilde{\tau} = t, \theta = H]}{\P[B_i = 1 \given \tilde{\tau} = t, \theta = L]} &= \frac{\P[B_i = 1 \given \tau_i \neq 0, \theta = H]}{\P[B_i = 1 \given \tau_i \neq 0, \theta = L]}\\
    &= \frac{\P[B_i = 1 \given \theta = H]}{\P[B_i = 1 \given \theta = L]} \cdot \frac{\P[\tau_i \neq 0 \given \theta = L]}{\P[\tau_i \neq 0 \given \theta = H]} \cdot
\end{align*}
Since this is bounded between $\frac{1}{\a^2}$ and $\a^2$, it follows that $\P[\theta = H \given \tilde{\tau} = t, B_i = 1] \in [\frac{1}{\a^2 + 1}, 1 - \frac{1}{\a^2 + 1}]$. Hence, by Lemma~\ref{lemma:product-signal},
\begin{align*}
    \I(\indic(\tau_0 = \tilde{\tau}) \given \tilde{\tau}) \leq 2 \cdot \frac{\log(\frac{1}{\a^2 + 1})}{\log(1 - \frac{1}{\a^2 + 1})} \leq 4 \a^2 \log(\a^2 + 1).
\end{align*}

It follows that $\I(\tau_0) \leq \log(\a) + 4 \a^2 \log(\a^2 + 1)$.

\end{proof}

\begin{proposition}
\label{prop:rt-extra-information-one-child}
There is a $D$ such that for any $T$ be a rooted tree with $k > 1$ ends and any associated adoption process, if the root has one child, then $\I(\tau_r) \leq \log(\a) + D + \I(\tau_i)$, where $i$ is the first descendant of the root with more than one child.
\end{proposition}

\begin{proof}
Let $j$ be the distance from the root to the first descendant with more than one child, and label each between the root and this child by the distance from the root.

Define $\tilde{\tau}$ as follows. If $\tau_i = 0$ for some $i < j$, then $\tilde{\tau}$ is the smallest such $i$. Otherwise, $\tilde{\tau} = \tau_j + j$. Observe that as in the proof of Proposition~\ref{prop:rt-one-end}, either $\tau_0 = \tilde{\tau}$ or $\tau_0 = \infty$. As before,
\begin{align*}
    \I(\tau_0) &\leq \I(\tilde{\tau}, \indic(\tau_0 = \tilde{\tau}))\\
    &= \I(\tilde{\tau}) + \I(\indic(\tau_0 = \tilde{\tau}) \given \tilde{\tau}).
\end{align*}

Note that as in the previous argument, $\I(\tilde{\tau}) \leq \max(\log(\a), \I(\tau_j))$. Moreover, by the same argument,
\begin{align*}
    \I(\indic(\tau_0 = \tilde{\tau}) \given \tilde{\tau}) \leq 4 \a^2 \log(\a^2 + 1).
\end{align*}
It follows that
\begin{align*}
    \I(\tau_0) &\leq \max(\log(\a), \I(\tau_j)) + 4 \a^2 \log(\a^2 + 1)\\
    &\leq \I(\tau_j) + \log(\a) + 4 \a^2 \log(\a^2 + 1).
\end{align*}

\end{proof}

Fix $c_0$ as in the first proposition and $D$ as in the second proposition. Define $c_1 = \max(c_0, \log(\a), D)$ and $C_1 = c_1$. For $k > 1$, define $c_k = (k+1) \cdot C_{k-1}$ and $C_k = c_k + \log(\a) + D$.

\begin{proposition}
For any rooted tree $T$ with $k$ ends and any associated adoption process, $\I(\tau_r) \leq C_k$.
\end{proposition}

\begin{proof}
We will prove by induction that if the root has more than one child, then $\I(\tau_r) \leq c_k$, and if the root has one child then $\I(\tau_r) \leq C_k$. Since $c_k < C_k$, the result then follows.

First, observe that if $k=1$, then the claim follows from Proposition~\ref{prop:rt-one-end}. So let $k > 1$, and assume the claim holds for $k' < k$.

First, observe that if the root has more than one child, then
\begin{align*}
    \I(\tau_r) \leq \I(s_r) + \sum_{i \in N_T(r)}{\I(\tau_i)}.
\end{align*}
Let $k_i$ be the number of ends of the subtree with root $i$. By assumption,
\begin{align*}
    \I(\tau_i) \leq C_{k_i}.
\end{align*}
Hence, it follows that
\begin{align*}
    \I(\tau_r) &\leq \log(\a) + \sum_{i \in N_T(r)}{C_{k_i}}\\
    &\leq \log(\a) + |N_T(r)| \cdot C_{k-1}\\
    &\leq \log(\a) + k \cdot C_{k-1}\\
    &\leq (k+1) \cdot C_{k-1}\\
    &= c_k.
\end{align*}
Now, suppose the root has one child, and let $i$ be the first descendant of $r$ that has more than one child. Since the subtree with root $i$ has $k$ ends and $i$ has more than one child, it follows that $\I(\tau_i) \leq c_k$. Hence,
\begin{align*}
    \I(\tau_r) &\leq \I(s_r) + \I(\tau_i) + D\\
    &\leq \log(\a) + D + c_k\\
    &= C_k.
\end{align*}

\end{proof}

Given a strategy profile, observe that the adoption times $\tau_i$ are functions of the private signals $s_i$. Given a pair of agents $i$ and $j$, we define the auxiliary random variable $\tau_j^i$ to be the adoption time of agent $j$ under the strategy profile where all agents but $i$ use the same strategy and agent $i$ uses the strategy under which they never adopt.

\begin{proof}[Proof of Theorem~\ref{thm:one-dimensional}]
Fix a one dimensional network $G$ and an agent $i$. Let $T$ be the rooted tree with $i$ as root and $j'$ a child of $j$ if $j$ is adjacent to $j'$ and the unique path from $i$ to $j'$ in $G$ contains $j$.

Given an equilibrium $\tau$, define the adoption process associated to $T$ as follows. The adoption time associated $i$ is $\tau_i$. For all other agents $j'$, the adoption time associated to $j'$ is $\tau_{j'}^{j}$, where $j$ is the parent of $j'$ in $T$.

Observe that $T$ has at most $2 M(G) + 2$ ends. It follows that $\I(\tau_i) \leq C_{2 M(G) + 2}$. Hence, by the bounding lemma,
\begin{align*}
    p_i \leq 1 - \frac{1}{2} e^{-C_{2 M(G) + 2} - 3}.
\end{align*}
\end{proof}

\section{Outside observer learns from period $0$ decisions}

Given a strategy profile $\tau$, denote by $p^O = \P[\theta = H \given (\indic(\tau_i = 0))_{i \in N}]$ the belief of an outside observer after seeing the agents' period $0$ decisions. We will say that an outside observer learns immediately from $\tau$ if
\begin{align*}
    \P[p^O = 1 \given \theta = H] = \P[p^O = 0 \given \theta = L] = 1.
\end{align*}

\begin{proposition}
\label{prop:outside-observer-learns-immediately}
If $G$ is an infinite undirected tree of bounded degree, then an outside observer learns immediately from any equilibrium strategy profile $\tau$ on $G$.
\end{proposition}

We will use the following result, which says that an outside observer learns immediately whenever there are infinitely many agents whose period $0$ adoption thresholds are uniformly bounded away from $b$.

\begin{lemma}
\label{lem:learn-immediately-if-uniformly-adopt-immediately}
Let $G$ be an infinite network and $\tau$ an equilibrium strategy profile. If there is an $\eta > 0$ such that $\P[\tau_i = 0 \given \theta = H] \geq \eta$ for infinitely many agents $i$, then an outside observer learns immediately from $\tau$.
\end{lemma}

\begin{proof}
Let
\begin{align*}
    \eps = \min\left( \frac{\eta}{\alpha}, \, \P[\pi_i < \frac12 \given \theta = H], \, \frac{\P[\pi_i \geq \frac12 \given \theta = H]}{\P[\pi_i \geq \frac12 \given \theta = L]} - 1\right).
\end{align*}
Observe that
\begin{align*}
    \P[\tau_i = 0 \given \theta = L] \geq \frac1\alpha \cdot \P[\tau_i = 0 \given \theta = H] \geq \frac1\alpha \cdot \eta \geq \eps,
\end{align*}
\begin{align*}
    \P[\tau_i > 0 \given \theta = H] \geq \P[\pi_i < \frac12 \given \theta = H] \geq \eps,
\end{align*}
and
\begin{align*}
    \frac{\P[\tau_i = 0 \given \theta = H]}{\P[\tau_i = 0 \given \theta = L]} \geq \frac{\P[\pi_i \geq \frac12 \given \theta = H]}{\P[\pi_i \geq \frac12 \given \theta = L]} \geq 1 + \eps,
\end{align*}
where the first inequality follows from the definition of $\alpha$ and the second and third follow from the fact that agents use threshold strategies in equilibrium. Hence, if $\P[\tau_i = 0 \given \theta = H] \geq \eta$, then
\begin{align*}
    \eps \leq \P[\tau_i = 0 \given \theta = L] = 1 - \P[\tau_i > 0 \given \theta = L] \leq 1 - \P[\tau_i > 0 \given \theta = H] \leq 1 - \eps,
\end{align*}
\begin{align*}
    \eps \leq \P[\tau_i = 0 \given \theta = L] \leq \P[\tau_i = 0 \given \theta = H] = 1 - \P[\tau_i > 0 \given \theta = H] \leq 1 - \eps,
\end{align*}
and
\begin{align*}
    \P[\tau_i = 0 \given \theta = H] \geq (1 + \eps) \cdot \P[\tau_i = 0 \given \theta = H].
\end{align*}

Now, fix $q \in (\frac12, 1)$, and let $m$ be as in Proposition~\ref{prop:seeing-many-nbrs-informative}, and let $S$ be any finite subset of agents such that $|S| \geq m$ and $\P[\tau_i = 0 \given \theta = H] \geq \eta$ for every agent $i \in S$. Denote by $p^S$ an outside observer's posterior after seeing the period $0$ adoption decisions of agents in $S$. Then
\begin{align*}
    \P[p^S \geq q \given \theta = H] \geq q
\end{align*}
and
\begin{align*}
    \P[p^S \leq 1 - q \given \theta = L] \geq q.
\end{align*}
Now, $\E[p^O \given p^S] = p^S$, so by Markov's inequality, if $p^S \geq q$, then
\begin{align*}
    \P[1 - p^O \geq \sqrt{1-q} \given p^S] \leq \frac{1 - p^S}{\sqrt{1 - q}} \leq \sqrt{1 - q},
\end{align*}
and if $p^S \leq 1 - q$, then
\begin{align*}
    \P[p^O \leq \geq \sqrt{1-q} \given p^S] \leq \frac{p^S}{\sqrt{1-q}} \leq \sqrt{1-q}.
\end{align*}
Thus, if $\max(p^S, 1 - p^S) \geq q$, then
\begin{align*}
    \P[\max(p^O, 1 - p^O) \leq 1 - \sqrt{1 - q} \given p^S] \leq \sqrt{1 - q}.
\end{align*}
Now,
\begin{align*}
    \P[\max(p^S, 1 - p^S) < q \given \theta = H] \leq \P[p^S < q \given \theta = H] \leq 1 - q
\end{align*}
and
\begin{align*}
    \P[\max(p^S, 1 - p^S) < q \given \theta = L] \leq \P[p^S > 1 - q \given \theta = L] \leq 1 - q,
\end{align*}
so $\P[\max(p^S, 1 - p^S) < q] \leq 1 - q$. It follows that $\P[\max(p^O, 1 - p^O) \geq q] \geq (1 - q)^{\frac32}$.

Since $q$ was arbitrary, it follows that $\P[\max(p^O, 1 - p^O) = 1] = 1$, and since $\P[p^O = 1 \given \theta = L] = \P[p^O = 0 \given \theta = H] = 0$, the result then follows.
\end{proof}

\begin{proof}[Proof of Proposition~\ref{prop:outside-observer-learns-immediately}]
Fix $\d \in (0, 1)$, and let $\tau$ be an equilibrium for the discount factor $\d$ on $G$. Let $d$ be the maximum degree of any agent in $G$. 

Let $\rho \in (0, 1)$ and $T \geq 1$ such that $\frac{1}{2 - \rho \cdot \d^T} < b$. Observe that for any agent $i$, if no agent within distance $T$ of $i$ adopts in period $0$, then by Proposition~\ref{prop:spontaneous}, agent $i$ does not adopt before period $T$. It follows that if the probability that no agent within distance $T$ of agent $i$ (other than $i$) adopts in period $0$ in the high state is at least $\rho$, then agent $i$'s expected utility from not adopting in period $0$ is less than $\pi_i \cdot \rho \cdot \d^T$. Hence, if $\pi_i \geq \frac{1}{2 - \rho \cdot \d^T}$, then agent $i$ must adopt in period $0$. Let
\begin{align*}
    q = \P[\pi_i \geq \frac{1}{2 - \rho \cdot \d^T} \given \theta = H]
\end{align*}
be the probability than an agent's private belief is above this threshold in the high state.

Denote by $A_i$ the event that at least one agent within a distance $T$ of agent $i$ (including agent $i$) adopts in period $0$. It follows from above that $\P[A_i^c \given \theta = H] \leq \max(\rho, 1-q)$. Now, since there are at most $d^{T+1}$ agents within a distance $T$ of agent $i$, it follows that there must be at least one agent $j$ within distance $T$ of agent $i$ such that $\P[\tau_j > 0 \given \theta = H] \leq \max(\rho, 1-q)^{\frac{1}{d^{T+1}}}$, and hence $\P[\tau_j = 0 \given \theta = H] \geq 1 - \max(\rho, 1 - q)^{\frac{1}{d^{T+1}}}$.

Finally, let $(i_n)$ be an infinite sequence of agents such that $d_G(i_m, i_n) > 2T$ for each $m, n$, and let $(j_n)$ be a sequence of agents such that for each $n$, $d_G(i_n, j_n) \leq T$ and $\P[\tau_{j_n} = 0 \given \theta = H] \geq 1 - \max(\rho, 1 - q)^{\frac{1}{d^{T+1}}}$. The result then follows from Lemma~\ref{lem:learn-immediately-if-uniformly-adopt-immediately}.
\end{proof}

\section{Proof of Theorem~\ref{thm:trees}}

\subsection{Auxiliary Continuous Time Model}

In this section we describe a continuous time decision making model that will be a useful tool for proving Theorem~\ref{thm:trees}.

The initial setup is as in our main model: there is a state $\theta \in \{H,L\}$ distributed uniformly. Private signals are denoted $s_i$ and take values in some measurable $S_i$. Private beliefs are $\pi_i = \P[\theta=H\given s_i]$, $\pi_i$ is almost surely in $[a,b]$ for some $0 < a < b < 1$, and the distribution of $\pi_i$ is not a point mass at the prior $1/2$. Agents can make irreversible adoption decisions. Not adopting yields a flow utility of $0$. Adopting yields a flow utility of $2$ if the state is high, and $-2$ if the state is low.


Hereon, the auxiliary model deviates from our main model. Time is continuous and takes values in $[0,1]$. The total utility from adopting at time $t \in [0,1]$ is $(1-t)$ if the state is high and $-(1-t)$ if the state is low. Note that adopting at time $1$ yields utility $0$ regardless of the state. 

The relation to our main model is as follows: for a given discount factor $\d<1$, the discrete time $t$ of our main model is reparameterized by $t \mapsto  1-\delta^t$. That is, adopting in discrete $t \in \{0,1,2,\ldots\}$, corresponds to adopting in continuous time in $\{0, 1-\delta, 1-\delta^2,\ldots\} \subset [0,1]$. We choose to discount by $1-t$ in the continuous time model precisely because the reparametrization $t \mapsto  1-\delta^t$ transforms exponential discounting by $\delta^t$ to discounting by $1-t$.

There is only one agent, the root agent or agent $0$. In addition to its private signal $s_0$ which it observes at time $t=0$, the root agent observes two random variables $\tau_1,\tau_2$, taking values in $[0,1]$. We think of these random variables as the adoption times of the root's children. We assume that they satisfy
\begin{align}
\label{info-process}
    \P[\tau_i \in T\given \theta=H ] \geq \P[\tau_i \in T\given \theta=L ] \text{ for all measurable } T \subset [0,1).
\end{align}
Since the continuous time $t=1$ corresponds to time infinity in the discrete time model, we think of $\tau_i=1$ as non-adoption. Hence, this condition is a monotonicity condition stating that adoption (at any time strictly before $t=1$) is more likely in the high state than the low state. As we shall see, this is satisfied by an optimizing agent. However, in this auxiliary model the children will have exogenously fixed behavior captured by $\tau_1$ and $\tau_2$. We assume that the joint distributions of $(\tau_1,\theta)$ and $(\tau_2,\theta)$ coincide and that $\tau_1$ and $\tau_2$ are independent conditional on $\theta$. This will allow us to think of the two children as participating in a symmetric equilibrium. 

Denote by $\mu$ the joint distribution of $(\tau_1,\theta)$. Since $\tau_1$ takes values in $[0,1]$ and $\theta$ takes values in $\{H,L\}$, $\mu$ is an element of $\Delta([0,1] \times \{H,L\})$. Note that this set of probability measures is compact in the weak topology, since $[0,1] \times \{H,L\}$ is compact. Denote by $\mu_H,\mu_L \in \Delta([0,1])$ the distributions of $\tau_1$ conditioned on $\theta$. Since in this auxiliary model we think of adopting at time $t=1$ as not adopting, the event that child $1$ is eventually correct is the event $[0,1) \times \{H\} \cup \{(1,L)\} \subset [0,1]\times \{H,L\}$, and the probability that it is eventually correct is $\frac{1}{2}\mu_H([0,1))+\frac{1}{2}\mu_L(\{1\})$. The condition of \eqref{info-process} can be written in terms of $\mu$ as
\begin{align}
\label{info-process-mu}
    \mu_H(T) \geq \mu_L(T) \text{ for all measurable } T \subset [0,1).
\end{align}

The expected discounted utility of each of the children is given by
\begin{align*}
    u(\mu) = \int_0^1(1-t)\,\dd\mu_H(t) - \int_0^1(1-t)\,\dd\mu_L(t).
\end{align*}
Note that the map $u \colon \Delta([0,1]\times \{H,L\}) \to [0,1]$ is continuous. 

We represent the strategy of the root agent by
a map $a \colon [0,1]^3 \to [0,1]$. This map calculates the root agent's adoption time, given its private belief and the adoption times of the children:
\begin{align*}
    \tau_0 = a(\pi_0,\tau_1,\tau_2).
\end{align*}
Given a strategy $a$ and the distribution $\mu$ describing the children's behavior, we denote by $w_\mu(a)$ the root agent's expected utility. Formally, if denote by $\nu_L$ and $\nu_H$ the conditional distributions of the private belief $\pi_0$, then
\begin{align*}
    w_\mu(a) = \int (1-a(\pi,t_1,t_2))\,\dd\nu_H(\pi)\dd\mu_H(t_1)\dd\mu_H(t_2) -  \int (1-a(\pi,t_1,t_2))\,\dd\nu_L(\pi)\dd\mu_L(t_1)\dd\mu_L(t_2).
\end{align*}

As explained above, we consider two families of strategies, each indexed by $r \in [0,1]$:
\begin{align*}
    a_{1,r}(\pi_0,\tau_1,\tau_2) &= 
    \begin{cases}
        \tau_1 &\text{if } \tau_1 > r\\
        \max\{\tau_2,r\} &\text{if } \tau_1 \leq r
    \end{cases}\\
    a_{2,r}(\pi_0,\tau_1,\tau_2) &= \begin{cases}
        r&\text{if } \tau_1  >r, \tau_2 \leq r, \pi_0>1/2\\
        \tau_1&\text{otherwise}.
    \end{cases}
\end{align*}
Denote this family of strategies by $\mathcal{A} = \{a_{i,r}\,:\, i \in \{1,2\}, r\in[0,1]\}$. Our next result shows that if the children have positive expected utility, and if they are not always eventually correct, then the root agent has a strategy in $\mathcal{A}$ that yields higher expected utility than the children. Note that both conditions are necessary for this conclusion.

Define 
\begin{align*}
    \eta(\mu) = \min\left\{u(\mu), \frac12 \mu_H\big(\{1\}\big) + \frac12 \mu_L\big([0, 1)\big)\right\}
\end{align*}
Note that the second term is a child's probability of not being eventually correct.

\begin{proposition}
    If $\eta(\mu) > 0$, then there exists a strategy $a \in \mathcal{A}$ such that $w_\mu(a) > u(\mu)$.
\end{proposition}
\begin{proof}
First, we claim that we can assume  that $(0,H)\in \mathrm{supp}(\mu)$. Otherwise, we can replace $0$ and $\varepsilon $ in the proof with $r$ and $r+\varepsilon $, where $r>0$ is the smallest number such that $(r, H)\in \mathrm{supp}(\mu)$; the same proof still goes through. As above, we denote by $\nu_H$ and $\nu_L$ the conditional distributions of $\pi_0$. 


We consider a number of cases. 

\paragraph{Case 1.} With positive probability $\tau_1=0$, so that in particular $\mu_H(0) > 0$. We break this into the following two subcases.

\paragraph{Case 1.1.} $\mu _{H}(0)=\mu _{L}(0)$. Denote by $c = (\mu_H(0)+\mu_L(0))/2$ the probability that $\tau_1=0$. In this case, conditioned on $\tau_1=0$, both states are equally likely, and hence the conditional expected utility of the child is zero. Hence conditioned on $\tau_1>0$, the conditional expected utility of the child is $u(\mu)/(1-c)$.

Consider the strategy $a_{1,0}$: follow child $1$ if $\tau_i>0$ and follow child $2$ otherwise.  Then this strategy will yield expected utility
\begin{align*}
    w_\mu(a_{1,0}) = (1-c)\frac{u(\mu)}{1-c} + c u(\mu) =(1+c)u(\mu) > u(\mu).
\end{align*}

\paragraph{Case 1.2.} $\mu _{H}(0)>\mu _{L}(0)$. The utility of the
strategy $a_{1,0}$ is
\begin{align*}
w_\mu(a_{1,0}) =& \mu _{H}(0)\int_{[0,1]}(1-t)\,\dd\mu
_{H}+\int_{[0,1]}(1-t)\,\dd\mu_{H}-\mu _{H}(0) \\
&-\left( \mu _{L}(0)\int_{[0,1]}(1-t)\,\dd\mu_{L}+\int_{[0,1]}(1-t)\,\dd\mu
_{L}-\mu _{L}(0)\right)  \\
=&u(\mu )-\left[ \mu _{H}(0)\left(1-\int_{[0,1]}(1-t)\,\dd\mu_{H}\right)-\mu
_{L}(0)\left(1-\int_{[0,1]}(1-t)\,\dd\mu _{L}\right)\right] .
\end{align*}

Denote by $p_H = \nu_H\left((1/2,1]\right)$ the probability that the root agent's private belief $\pi_0$ is larger than $1/2$ in the high state, and define $p_L$ analogously. Then the expected utility of the strategy $a_{2,0}$ is
\begin{align*}
w_\mu(a_{2,0}) =&p_H\cdot\mu _{H}(0)+\left[ 1-p_H\cdot\mu _{H}(0)\right] \int_{[0,1]}(1-t)\,\dd\mu _{H} \\
&-\left( p_L\cdot\mu _{L}(0)+\left[ 1-p_L\cdot\mu _{L}(0)%
\right] \int_{[0,1]}(1-t)\,\dd\mu_{L}\right)  \\
=&u(\mu )+p_H\cdot\mu _{H}(0)\left[ 1-\int_{[0,1]}(1-t)\,\dd\mu_{H}%
\right] -p_L\cdot\mu _{L}(0)\left[ 1-\int_{[0,1]}(1-t)\,\dd\mu_{L}%
\right] .
\end{align*}%
If $\mu _{L}(0)\left(1-\int_{0}^{1}(1-t)\,\dd\mu _{L}\right)>\mu
_{H}(0)\left(1-\int_{[0,1]}(1-t)\,\dd\mu_{H}\right)$, then $a_{1,0}$ will make the root agent
better off than her children. Otherwise, we have%
\begin{equation*}
\mu _{H}(0)\left[ 1-\int_{[0,1]}(1-t)\,\dd\mu _{H}\right] \geq \mu _{L}(0)\left[
1-\int_{[0,1]}(1-t)\,\dd\mu_{L}\right] 
\end{equation*}%
so that%
\begin{equation*}
p_H\cdot\mu _{H}(0)\left[ 1-\int_{[0,1]}(1-t)\,\dd\mu _{H}\right] >p_L\cdot\mu _{L}(0)\left[ 1-\int_{[0,1]}(1-t)\,\dd\mu_{L}\right] 
\end{equation*}
since $p_H > p_L$, as the root agent's signal is informative. Hence $a_{2,0}$ will make the root agent better off than her children. 

\paragraph{Case 2.} Having considered these cases, we can assume henceforth that $\tau_1=0$ with probability zero. Hence $\mu_H(0)=0$ and $\lim_{\varepsilon \downarrow 0}\mu _{H}([0,\varepsilon ])=0$. Let $\varepsilon >0$.

We first analyze the strategy $a_{1,\varepsilon }$. We have 
\begin{align*}
w_\mu(a_{1,\varepsilon }) =&\mu_{H}([0,\varepsilon])\int_{(\varepsilon,1]}(1-t)\,\dd\mu _{H}+\int_{(\varepsilon,1]}(1-t)\,\dd\mu
_{H}  \\
&-\mu _{L}[0,\varepsilon ]\int_{(\varepsilon,1]}(1-t)\,\dd\mu
_{L}-\int_{(\varepsilon,1]}(1-t)\,\dd\mu _{L} \\
&+(1-\varepsilon)\left(\mu _{H}([0,\varepsilon])^{2}-\mu
_{L}([0,\varepsilon])^{2}\right)  \\
\geq &u(\mu )-\int_{[0,\varepsilon] }(1-t)\,\dd\mu
_{H}+\int_{[0,\varepsilon] }(1-t)\,\dd\mu _{L}\\
&+\mu _{H}([0,\varepsilon
])\int_{(\varepsilon,1]}(1-t)\,\dd\mu _{H}-\mu _{L}([0,\varepsilon
])\int_{(\varepsilon,1]}(1-t)\,\dd\mu _{L}   \\
=&u(\mu )-\mu _{H}([0,\varepsilon ])\left[ 1-\int_{(\varepsilon,1]}(1-t)\,\dd\mu _{H}\right] +\mu _{L}([0,\varepsilon ])\left[
1-\int_{(\varepsilon,1]}(1-t)\,\dd\mu _{L}\right]\\
&+\left( \int_{[0,\varepsilon]
}t\,\dd\mu _{H}-\int_{[0,\varepsilon] }t\,\dd\mu _{L}\right)   \notag \\
\geq &u(\mu )-\mu _{H}([0,\varepsilon ])\left[ 1-\int_{(\varepsilon,1]}(1-t)\,\dd\mu _{H}\right] +\mu _{L}([0,\varepsilon ])\left[
1-\int_{(\varepsilon,1]}(1-t)\,\dd\mu _{L}\right] .  
\end{align*}
We proceed to the second strategy $a_{2,\varepsilon }$:%
\begin{align*}
w_\mu(a_{2,\varepsilon }) =&p_H\cdot \mu
_{H}([0,\varepsilon ])\mu _{H}([0,1-\varepsilon ])(1-\varepsilon
)+\int_{[0,\varepsilon ]}(1-t)\,\dd\mu _{H}\\
&+[1-p_H\cdot \mu
_{H}([0,\varepsilon ])]\int_{(\varepsilon,1]}(1-t)\,\dd\mu _{H} \\
&-p_L\cdot \mu _{L}([0,\varepsilon ])\mu _{L}([0,1-\varepsilon
])(1-\varepsilon )+\int_{[0,\varepsilon] }(1-t)\,\dd\mu _{L}\\
&+[1-p_L\cdot \mu _{L}([0,\varepsilon ])]\int_{(\varepsilon,1]}(1-t)\,\dd\mu _{L} \\
=&u (\mu )+p_H\cdot \mu _{H}([0,\varepsilon ])\left( \mu
_{H}([0,1-\varepsilon ])(1-\varepsilon )-\int_{(\varepsilon,1]}(1-t)\,\dd\mu
_{H}\right)  \\
&-p_L\cdot \mu _{L}([0,\varepsilon ])\left( \mu
_{L}([0,1-\varepsilon ])(1-\varepsilon )-\int_{(\varepsilon,1]}(1-t)\,\dd\mu
_{L}\right) .
\end{align*}%
Note that 
\begin{equation*}
\lim_{\varepsilon \downarrow 0}\frac{1-\int_{(\varepsilon,1]}(1-t)\,\dd\mu _{H}%
}{1-\int_{(\varepsilon,1]}(1-t)\,\dd\mu _{L}}=\frac{1-\int_{[0,1]}(1-t)\,\dd\mu _{H}%
}{1-\int_{[0,1]}(1-t)\,\dd\mu _{L}}=\lim_{\varepsilon \downarrow 0}\frac{\mu
_{H}([0,1-\varepsilon ])(1-\varepsilon )-\int_{(\varepsilon,1]}(1-t)\,\dd\mu
_{H}}{\mu _{L}([0,1-\varepsilon ])(1-\varepsilon )-\int_{(\varepsilon,1]}(1-t)\,\dd\mu _{L}}.
\end{equation*}%
Since the root agent's signal is informative,  $p_H >p_L$, and so whenever $\varepsilon $ is small enough, we must have one of the following is true: either
\begin{align*}
\lefteqn{p_H\cdot \mu _{H}([0,\varepsilon ])\left( \mu _{H}([0,1-\varepsilon
])(1-\varepsilon )-\int_{(\varepsilon,1]}(1-t)\,\dd\mu _{H}\right)}  \\
&\geq p_L\cdot \mu _{L}([0,\varepsilon ])\left( \mu
_{L}([0,1-\varepsilon ])(1-\varepsilon )-\int_{(\varepsilon,1]}(1-t)\,\dd\mu
_{L}\right),
\end{align*}
or
\begin{equation*}
\mu _{L}([0,\varepsilon ])\left[ 1-\int_{(\varepsilon,1]}(1-t)\,\dd\mu _{L}\right]
\geq \mu _{H}([0,\varepsilon ])\left[ 1-\int_{(\varepsilon,1]}(1-t)\,\dd\mu _{H}%
\right] .
\end{equation*}%
This translates to $w_\mu(a_{2,\varepsilon})>u(\mu )$ or $%
w_\mu(a_{1,\varepsilon })>u(\mu )$. 
\end{proof}

Define $\Psi(\mu) = \sup_{a \in \mathcal{A}}{w_{\mu}(a)}$. Note that $\Psi$ provides a lower bound on the root's maximum expected utility when the children's adoption times follow $\mu$. As the above Proposition shows, if $\eta(\mu) > 0$ then $\Psi(\mu) > u(\mu)$. However, we will need a stronger quantitative result. For $\eps \in (0, 1)$, let
\begin{align*}
    C_{\eps} = \inf_{\mu \, : \, \eta(\mu) \geq \eps}{\frac{\Psi(\mu)}{u(\mu)}} \cdot
\end{align*}
This captures the guaranteed improvement factor of the root's utility over the children's utility, given that the distribution of their adoption times $\mu$ satisfies $\eta(\mu) > \varepsilon$, i.e., that their expected discounted utility is at least $\varepsilon$ and their probability of not being eventually correct is at least $\varepsilon$.

\begin{proposition}
\label{prop:Ceps}
$C_{\eps} > 1$ for all $\eps \in (0, 1)$.
\end{proposition}
We will require a few lemmas to prove this claim.

For any distributions $\mu, \mu'$, we denote by $D_T(\mu, \mu')$ the transportation metric given by the minimum of $\E[|\tau-\tau'|]$ taken over all pairs of random variables $\tau,\tau'$ distributed $\mu,\mu'$ respectively. We recall that this distance metrizes the weak topology, i.e., $\lim_n \mu_n = \mu$ if and only if $\lim_n D_T(\mu_n,\mu) = 0$.
\begin{lemma}
\label{lem:DT}
Fix $\mu, \mu'$. For any $\eps \in (0, 1)$, $i \in \{1, 2\}$, and $r \in [0, 1]$, if $D_T(\mu, \mu') < \eps^2$ then
\begin{align*}
    |w_{\mu}(a_{i, r}) - w_{\mu'}(a_{i, r})| \leq 2 \cdot \mu\big([r-\eps, r+\eps] \times \{L, H\}\big) + 3 \cdot \eps.
\end{align*}
\end{lemma}

\begin{proof}
Let $\tau_1, \tau_2$ each have law $\mu$ and $\tau_1', \tau_2'$ each have law $\mu'$, with $\E[|\tau_j - \tau_j'|] < \eps^2$ for $j = 1, 2$. Let $\tau = a_{i, r}(\pi_0, \tau_1, \tau_2)$ and $\tau' = a_{i, r}(\pi_0, \tau_1', \tau_2')$. Observe that
\begin{align*}
    w_{\mu}(a_{i, r}) - w_{\mu'}(a_{i, r}) &= \E[(\indic(\theta = H) - \indic(\theta = L)) \cdot (1 - \tau)] - \E[(\indic(\theta = H) - \indic(\theta = L)) \cdot (1 - \tau')]\\
    &= \E[(\indic(\theta = H) - \indic(\theta = L)) \cdot (\tau' - \tau)],
\end{align*}
so
\begin{align*}
    |w_{\mu}(a_{i, r}) - w_{\mu'}(a_{i, r})| \leq \E[|(\indic(\theta = H) - \indic(\theta = L)) \cdot (\tau' - \tau)|] = \E[|\tau - \tau'|].
\end{align*}
Let $A$ be the event that $|\tau_1 - \tau_1'| < \eps$, $|\tau_2 - \tau_2'| < \eps$, $|\tau_1 - r| > \eps$, and $|\tau_2 - r| > \eps$. As we now show, if $A$ obtains, then $|\tau - \tau'| < \eps$. We consider two cases, depending on whether $a_{i,r}=a_{1,r}$ or $a_{i,r}=a_{2,r}$.
\paragraph{Case 1.} $i = 1$\\
If $\tau_1 > r$, then $\tau_1' > r$, so $|\tau - \tau'| = |\tau_1 - \tau_1'| < \eps$. If $\tau_1 < r$ and $\tau_2 < r$, then $|\tau - \tau'| = |\tau_2 - \tau_2'| < \eps$. If $\tau_1 < r$ and $\tau_2 > r$, then $|\tau - \tau'| = |r - r| = 0$.
\paragraph{Case 2.} $i = 2$\\
Observe that $\tau_1 > r$ and $\tau_2 \leq r$ if and only if $\tau_1' > r$ and $\tau_2' \leq r$. Hence, either $|\tau - \tau'| = |\tau_1 - \tau_1'| < \eps$ or $|\tau - \tau'| = |r - r| = 0$.\\

\noindent Hence, since $|\tau - \tau'| \leq 1$ almost surely, it follows that
\begin{align*}
    |w_{\mu}(a_{i, r}) - w_{\mu'}(a_{i, r})| \leq \P[A] \cdot \eps + (1 - \P[A]) \cdot 1 \leq \eps + 1 - \P[A].
\end{align*}
Observe that by the union bound,
\begin{align*}
    1 - \P[A] \leq \P[|\tau_1 - \tau_1'| \geq \eps] + \P[|\tau_2 - \tau_2'| \geq \eps] + \P[|\tau_1 - r| \leq \eps] + \P[|\tau_2 - r| \leq \eps].
\end{align*}
Now, $\P[|\tau_j - r| \leq \eps] = \mu\big((r - \eps, r+\eps) \times \{L, H\}\big)$, and by Markov's inequality, $$
    \P[|\tau_j - \tau_j'| \geq \eps] \leq \frac{\E[|\tau_j - \tau|]}{\eps} \leq \frac{\eps^2}{\eps} = \eps.
$$
Hence,
\begin{align*}
    1 - \P[A] \leq 2 \eps + 2 \mu\big((r - \eps, r+\eps) \times \{L, H\}\big)
\end{align*}
and the result follows.

\end{proof}

\begin{lemma}
\label{lem:rj}
Fix $\mu$. For any $\g > 0$, there exists $i \in \{1, 2\}$, $r \in [0, 1]$, and $\eps > 0$ such that $\mu(\{r\} \times \{L, H\}) = 0$, $\mu\big([r - \eps, r + \eps] \times \{L, H\}\big) < \g$, and $w_{\mu}(a_{i, r}) > \Psi(\mu) - \g$.
\end{lemma}

\begin{proof}
Observe first that $w_{\mu}(a_{1, 1}) = 0$ and $w_{\mu}(a_{2, 1}) = u(\mu)$. Since $\Psi(\mu) > u(\mu)$, it follows that there is some $i \in \{1, 2\}$ and $r \in [0, 1)$ such that $w_{\mu}(a_{i, r}) > \Psi(\mu) - \frac12 \cdot \g$.

Now, let $r_j$ be a descending sequence such that $\mu(\{r_j\} \times \{L, H\}) = 0$ and $\lim r_j = r$. Now, if $\tau_1, \tau_2 \notin (r, r_j]$, then $a_{i, r_j}(\pi_0, \tau_1, \tau_2) = a_{i, r}(\pi_0, \tau_1, \tau_2)$, i.e., the action taken under strategy $a_{i,r_j}$ and $a_{i,r}$ is the same. In particular, since $\lim_j\mu((r, r_j] \times \{L, H\}) = 0$, this happens with probability tending to one, and so, because payoffs are bounded, it follows that $\lim_j w_{\mu}(a_{i, r_j}) = w_{\mu}(a_{i, r})$. 

Choose $r_j$ such that $w_{\mu}(a_{i, r_j}) > w_{\mu}(a_{i, r}) - \frac12 \cdot \g$. Then $w_{\mu}(a_{i, r_j}) > \Psi(\mu) - \g$. Since $\lim_{\eps \to 0} \mu\big([r_j - \eps, r_j + \eps] \times \{L, H\}\big) = 0$, the result then follows.
\end{proof}

Given these lemmas we are ready to prove the proposition
\begin{proof}[Proof of Proposition~\ref{prop:Ceps}]
Fix $\eps \in (0, 1)$, and let $\mathcal{C} \subseteq \Delta([0, 1] \times \{H, L\})$ be the set of all measures $\mu$ such that $\mu$ satisfies the monotonicity condition \eqref{info-process-mu} and $\eta(\mu) \geq \eps$. Observe that since $u$ is continuous with respect to the weak topology, $\eta$ is continuous, and since \eqref{info-process-mu} is a closed condition, it follows that $\mathcal{C}$ is closed. Moreover, since $\Delta([0, 1] \times \{H, L\})$ is compact, it follows that $\mathcal{C}$ is compact.

Now, $C_{\eps} = \inf_{\mu \in \mathcal{C}}{\frac{\Psi(\mu)}{u(\mu)}}$ by definition, and since $\Psi(\mu) > u(\mu)$ for all $\mu \in \mathcal{C}$, it is sufficient to show that $\frac{\Psi(\mu)}{u(\mu)}$ achieves its infimum on $\mathcal{C}$, and for this it is sufficient to show that $\frac{\Psi(\mu)}{u(\mu)}$ is lower semi-continuous. Moreover, since $u(\mu)$ is continuous, it is sufficient to show that $\Psi(\mu)$ is lower semi-continuous. 

So suppose $\lim_n \mu^n = \mu$. Fix $\g > 0$. By Lemma~\ref{lem:rj}, there exist $i \in \{1, 2\}$ and $r \in [0, 1]$ such that for all sufficiently small $\zeta > 0$, $\mu(\{r\} \times \{L, H\}) = 0$, $\mu\big([r - \zeta, r + \zeta] \times \{L, H\}\big) < \g$, and $w_{\mu}(a_{i, r}) > \Psi(\mu) - \g$. By Lemma \ref{lem:DT}, it follows if $\zeta < \g$ is sufficiently small, then for all sufficiently large $n$,
\begin{align*}
    \Psi(\mu^n) &\geq w_{\mu^n}(a_{i, r})\\
    &\geq w_{\mu}(a_{i, r}) - 2 \cdot \g - 3 \cdot \zeta\\
    &\geq \Psi(\mu) - 6 \cdot \g.
\end{align*}
Hence, $\liminf \Psi(\mu^n) \geq \Psi(\mu) - 6 \cdot \g$. Since this holds for all $\g$, it follows that $\liminf \Psi(\mu^n) \geq \Psi(\mu)$. Hence, $\Psi$ is lower semi-continuous, and the result follows.
\end{proof}

We can now prove Theorem~\ref{thm:trees}.

\begin{proof}[Proof of Theorem~\ref{thm:trees}]

Fix $\bar{p} < 1$. Let $u_0$ be the expected utility from adopting in period $0$ if $\pi_i \geq \frac12$ and never adopting otherwise. Note that $u_0>0$, since signals are informative. Let $\eps = \min\{1-\bar{p},u_0\}$. Fix any discount factor $\d <1$ such that $C_{\eps} \cdot \d > 1$.

Suppose towards a contradiction that there is a symmetric equilibrium in which each agent's probability of being eventually correct is less than $\bar{p}$. Fix an agent $0$ and two of its children, $1$ and $2$. Let $\tilde{\tau}_i = 1 - \d^{\tau_i}$, and let $\mu$ be the distribution of $\tilde{\tau}_i$. Note that in equilibrium, $\P[\tau_i = t \given \theta = H] \geq \P[\tau_i = t \given \theta = L]$ by Lemma~\ref{lem:adoption-more-likely-in-high-state}, so $\mu$ satisfies the monotonicity condition \eqref{info-process-mu}. By the definition of $\tilde{\tau}$, the expected utilities of the children are exactly equal to the expected utilities $u(\mu)$ in the continuous time model.

Observe first that $u(\mu) \geq u_0$, since the strategy described above is always available and agents are maximizing expected utility. Moreover, since the probability of not being eventually correct is at least $1-\bar{p}$, $\eta(\mu) \geq \eps$. Hence, by the definition of $C_\eps$, 
$$
  \Psi(\mu) \geq C_{\eps} \cdot u(\mu) > \frac{1}{\d} \cdot u(\mu).
$$
In particular, there is some strategy $a_{i,r}$ in the continuous time model  such that $w_{\mu}(a_{i,r}) > \frac{1}{\d} \cdot u(\mu)$. Note that $\tilde{\tau}_i$ is supported on $\{1 - \d^n\}_{n \geq 0} \cup \{1\}$. Since no additional information is provided between $1 - \d^n$ and $1 - \d^{n+1}$, it follows that we can take $r$ to be in $\{1 - \d^n\}_{n \geq 0} \cup \{1\}$, in which case the root agent only takes actions in this set.

Finally, let $\tau_0 = \frac{\log(1 - a_{i, r}(\pi_0, \tilde{\tau}_1, \tilde{\tau}_2))}{\log(\d)} + 1$. Observe that $\tau_0$ is a valid strategy, since $\indic(\tau_0 = t)$ depends only on $s_0$ and $\indic(\tau_i = t')$ for $i = 1, 2$ and $t' < t$. Finally, denoting by $\nu$ the distribution of $a_{i, r}(\pi_0, \tilde{\tau}_1, \tilde{\tau}_2)$, observe that agent $0$'s expected utility is equal to $\d \cdot u(\nu) > u(\mu)$. Since every agent's expected utility is $u(\mu)$, it follows that agent $0$ is not best responding, contradicting the equilibrium assumption.

Thus, for all $\d > \frac{1}{C_{\eps}}$, every agent's probability of being eventually correct is at least $\bar{p}$ in every symmetric equilibrium.
\end{proof}

\section{Learning from period $0$ actions}

For several results, it is necessary to establish quantitative bounds on the information contained in an agent's period $0$ action. Denote by $X_j = \indic(\tau_j = 0)$ be the indicator of the event that agent $j$ adopts in period $0$, and denote
\begin{align*}
    \chi_j = \log \frac{\P[X_j \given \theta = H]}{\P[X_j \given \theta = L]} \cdot
\end{align*}

\begin{lemma}
\label{lem:seeing-one-nbr-informative}
For any $\eps > 0$ there exist $\rho, \rho' > 0$ such that if $\eps \leq \P[X_j = 1 \given \theta] \leq 1 - \eps$ for $\theta \in \{L, H\}$ and $\P[X_j = 1 \given \theta = H] \geq (1 + \eps) \cdot \P[X_j = 1 \given \theta = L]$, then 

$\E[\chi_j \given \theta = L] \leq -\rho$, $\E[\chi_j \given \theta = H] \geq \rho$, and $\Var[\chi \given \theta] \leq \rho'$.
\end{lemma}

Since the agents' period $0$ decisions are conditionally independent, this allows us to establish a quantitative bound on the information from observing many agents' period $0$ actions. For any finite set of agents $S$ denote $\chi_S = \sum_{j \in S} \chi_j$.

\begin{proposition}
\label{prop:seeing-many-nbrs-informative}
For any $\eps > 0$ and $q \in (\frac12, 1)$, there is an $m$ such that if $|S| \geq m$, $\eps \leq \P[X_j = 1 \given \theta] \leq 1 - \eps$ for $\theta \in \{L, H\}$ and $\P[X_j = 1 \given \theta = H] \geq (1 + \eps) \cdot \P[X_j = 1 \given \theta = L]$ for every $j \in S$, then
\begin{align*}
    \P[\chi_S \geq \log \frac{q}{1 - q} \given \theta = H] \geq q
\end{align*}
and
\begin{align*}
    \P[\chi_S \leq - \log \frac{q}{1-q} \given \theta = L] \geq q.
\end{align*}
\end{proposition}

\begin{proof}
Let $\rho, \rho'$ be as in Lemma~\ref{lem:seeing-one-nbr-informative}, and let $x_{\theta} = \E[\chi_S \given \theta]$ and $v_\theta = \Var[\chi_S \given \theta]$. Then $x_H \geq |S| \cdot \rho$, $x_L \leq -|S| \cdot \rho$, and $v_\theta \leq |S| \cdot \rho'$, so for any $t < |S| \cdot \rho$,
\begin{align*}
    \P[\chi_S < t \given \theta = H] &= \P[x_H - \chi > x_H - t \given \theta = H]\\
    &\leq \P[(x_H - \chi)^2 > (x_H - t)^2 \given \theta = H]\\
    &\leq \frac{v_H}{(x_H - t)^2}\\
    &\leq \frac{|S| \cdot \rho'}{(|S| \cdot \rho - t)^2},
\end{align*}
and by an analogous calculation, 
\begin{align*}
    \P[\chi_S > -t \given \theta = L] \leq \frac{|S| \cdot \rho'}{(|S| \cdot \rho - t)^2} \cdot
\end{align*}
Now, taking $t = \log \frac{q}{1 - q}$, $t < m \cdot \rho$ for all sufficiently large $m$ and 
\begin{align*}
    \lim_{m \to \infty}{\frac{m \cdot \rho'}{(m \cdot \rho - t)^2}} = 0.
\end{align*}
Hence, there is an $m$ such that if $|S| \geq m$ then $t \leq |S| \cdot \rho$ and 
\begin{align*}
    \frac{|S| \cdot \rho'}{(|S| \cdot \rho - t)^2} \leq q,
\end{align*}
the result then follows.
\end{proof}

\begin{proof}[Proof of Lemma~\ref{lem:seeing-one-nbr-informative}]
Let
\begin{align*}
    f_H(p, q) &= (1-p) \cdot \log \left( \frac{1-p}{1-q} \right) + p \cdot \log \left( \frac{p}{q} \right)\\
    f_L(p, q) &= (1-q) \cdot \log \left( \frac{1-p}{1-q} \right) + q \cdot \log \left( \frac{p}{q} \right)\\
    g_H(p, q) &= (1-p) \cdot \left( \log \left( \frac{1-p}{1-q} \right) - f_H(p, q) \right)^2 + p \cdot \left( \log \left( \frac{p}{q} \right) - f_H(p, q) \right)^2\\
    g_L(p, q) &= (1-q) \cdot \left( \log \left( \frac{1-p}{1-q} \right) - f_L(p, q) \right)^2 + q \cdot \left( \log \left( \frac{p}{q} \right) - f_L(p, q) \right)^2.
\end{align*}
Further, let $Z = \{(p, q) \, : \, p, q \in [\eps, 1 - \eps], \, p \geq (1 + \eps) \cdot q\}$. By Jensen's inequality, $f_H(p, q) > 0$ and $f_L(p, q) < 0$ for all $(p, q) \in Z$.
Since $f_L$, $f_H$, $g_L$, and $g_H$ are continuous and $Z$ is closed, it follows that there are $\rho, \rho' > 0$ such that $f_L(p, q) \leq - \rho$, $f_H(p, q) \geq \rho$, $g_L(p, q) \leq \rho'$, and $g_H(p, q) \leq \rho'$ for all $(p, q) \in Z$. The result then follows, since
\begin{align*}
    \E[\chi \given \theta = H] &= f_H(\P[X = 1 \given \theta = H], \P[X = 1 \given \theta = L])\\
    \E[\chi \given \theta = L] &= f_L(\P[X = 1 \given \theta = H], \P[X = 1 \given \theta = L])\\
    \Var[\chi \given \theta = H] &= g_H(\P[X = 1 \given \theta = H], \P[X = 1 \given \theta = L])\\
    \Var[\chi \given \theta = L] &= g_L(\P[X = 1 \given \theta = H], \P[X = 1 \given \theta = L]).
\end{align*}
\end{proof}


\end{document}